\documentclass[conference]{IEEEtran}
\IEEEoverridecommandlockouts

\usepackage{graphicx}
\usepackage{cite}
\usepackage{url}
\usepackage{color}
\usepackage{alltt}
\usepackage{syntax}
\usepackage{newfloat}
\usepackage{mdframed}
\usepackage{multicol}
\usepackage{amsmath}
\usepackage{tikz}
\usepackage{filecontents} 
\usepackage{graphicx}
\usepackage{epstopdf} 
\usepackage{xspace}
\usepackage{enumitem}
\usepackage{svg}
\usepackage{lscape}
\usepackage{longtable}
\usepackage{supertabular}

\usepackage[ruled,linesnumbered,noend]{algorithm2e}
\SetKw{Continue}{continue}
\SetKwInput{KwInput}{Input}
\SetKwInput{KwOutput}{Output}
\SetKwProg{Fn}{Function}{}{end}

\SetCommentSty{mycommfont}

\usepackage{adjustbox}
\usepackage{wrapfig}
\usepackage{graphbox}
\usepackage{rotating}
\usepackage{pdfpages}
\usepackage{mdwlist}
\usepackage{multirow}
\usepackage[T1]{fontenc}

\usepackage{subcaption}
\usepackage[labelfont=bf,skip=0pt]{caption}
\usepackage{hyperref} 
\hypersetup{
	citebordercolor={0 1 0},
	citecolor=blue,
	colorlinks=false,
	filebordercolor={0 .5 .5},
	filecolor=green,
	linkbordercolor={1 0 0},
	linkcolor=blue,
	menubordercolor={1 0 0},
	pageanchor=true,
	pagebackref=true,
	pdfborder={0 0 1},
	pdfpagelabels=true,
	pdftex,
	plainpages=false,
	urlbordercolor={0 1 1},
	urlcolor=blue,
}

\usepackage[capitalize,nameinlink]{cleveref}
\hypersetup{%
	bookmarksnumbered, bookmarksopen=true, bookmarksopenlevel=1,%
}
\crefname{figure}{Figure}{Figures}
\crefname{listing}{Query}{Queries}
\crefname{section}{Section}{Sections}
\crefname{table}{Table}{Tables}
\crefname{BNF}{Grammar}{Grammars}
\crefname{algorithm}{Algorithm}{Algorithms}
\crefname{equation}{Equation}{Equations}

\usepackage{listings}
\definecolor{mygreen}{rgb}{0,0.6,0}
\definecolor{mygray}{rgb}{0.5,0.5,0.5}

\newcommand{\incode}[1]{\lstinline{#1}}

\newcommand{\distance}{5pt}
\DeclareFloatingEnvironment[
fileext   = logr,
listname  = {List of Grammars},
name      = Grammar,
placement = !htbp
]{BNF}

\newcommand{\msim}{\raise.17ex\hbox{$\scriptstyle\sim$}}

\newcommand{\myparatight}[1]{\smallskip\noindent{\bf {#1}.}}

\newcommand{\eat}[1]{}
\newcommand{\eg}{e.g.,\xspace}
\newcommand{\ie}{i.e.,\xspace}
\newcommand{\aka}{a.k.a.,\xspace}

\newcommand{\tool}{\textsc{ThreatRaptor}\xspace}
\newcommand{\lang}{TBQL\xspace}
\newcommand{\cti}{OSCTI\xspace}

\def\BibTeX{{\rm B\kern-.05em{\sc i\kern-.025em b}\kern-.08em
    T\kern-.1667em\lower.7ex\hbox{E}\kern-.125emX}}

\begin{document}

\title{Enabling Efficient Cyber Threat Hunting With Cyber Threat Intelligence}

\author{
\IEEEauthorblockN{Peng Gao\IEEEauthorrefmark{1},
Fei Shao\IEEEauthorrefmark{2},
Xiaoyuan Liu\IEEEauthorrefmark{1},
Xusheng Xiao\IEEEauthorrefmark{2},
Zheng Qin\IEEEauthorrefmark{3},
Fengyuan Xu\IEEEauthorrefmark{3}\\
Prateek Mittal\IEEEauthorrefmark{4},
Sanjeev R. Kulkarni\IEEEauthorrefmark{4},
Dawn Song\IEEEauthorrefmark{1}}
\IEEEauthorblockA{\IEEEauthorrefmark{1}University of California, Berkeley    \IEEEauthorrefmark{2}Case Western Reserve University \\   \IEEEauthorrefmark{3}National Key Lab for Novel Software Technology, Nanjing University
\IEEEauthorrefmark{4}Princeton University \\
\IEEEauthorrefmark{1}\{penggao,xiaoyuanliu,dawnsong\}@berkeley.edu    \IEEEauthorrefmark{2}\{fxs128,xusheng.xiao\}@case.edu   \\ \IEEEauthorrefmark{3}\{qinzheng,fengyuan.xu\}@nju.edu.cn
\IEEEauthorrefmark{4}\{pmittal,kulkarni\}@princeton.edu
}
}

\maketitle

\begin{abstract}

Log-based 
cyber 
threat hunting has emerged as an important solution to counter sophisticated
attacks.
However, existing approaches require non-trivial efforts of manual query construction and have overlooked the rich external threat knowledge
provided by open-source Cyber Threat Intelligence (\cti).
To bridge the gap, 
we propose \tool, a system that facilitates
threat hunting in computer systems using \cti.
Built upon 
system auditing frameworks, \tool provides 
(1) an unsupervised, light-weight, and accurate NLP pipeline that extracts structured threat behaviors from unstructured \cti text,
(2) a concise and expressive domain-specific query language, \lang, to hunt for malicious system activities,
(3) a query synthesis mechanism that automatically synthesizes a \lang query for 
hunting,
and 
(4) an efficient query execution engine to search the big 
audit logging data.
Evaluations on a broad set of attack cases demonstrate the accuracy and efficiency of \tool in 
practical threat hunting.

\end{abstract}

\section{Introduction}
\label{sec:intro}

Recent cyber attacks have plagued many well-protected businesses
~\cite{target,equifax}.
These attacks often exploit multiple types of vulnerabilities to infiltrate into target systems in multiple stages, posing challenges
for effective countermeasures.
To counter these attacks, \emph{ubiquitous system auditing} has emerged as an important approach for monitoring system activities~\cite{backtracking,lee2013high,liu2018priotracker,milajerdi2019poirot}.
System auditing collects system-level auditing events about system calls from OS kernel as system audit logs.
The collected system audit logging data further enables approaches to hunt for cyber threats via query processing~\cite{gao2018aiql,gao2018saql,pasquier2018runtime}.

Cyber threat hunting 
is the process of proactively and iteratively searching for malicious actors and indicators in various logs,
which is critical to early-stage detection.
Despite numerous research outcomes~\cite{gao2018aiql,gao2018saql,pasquier2018runtime} and industry solutions~\cite{splunk-spl,elastic-siem},
existing approaches, however, require non-trivial efforts of manual query construction and have overlooked the rich external knowledge about threats
provided by open-source Cyber Threat Intelligence (\cti).
Hence, the current threat hunting process is labor-intensive and error-prone.

\begin{figure*}[t]
    \centering
    \includegraphics[width=.975\linewidth]{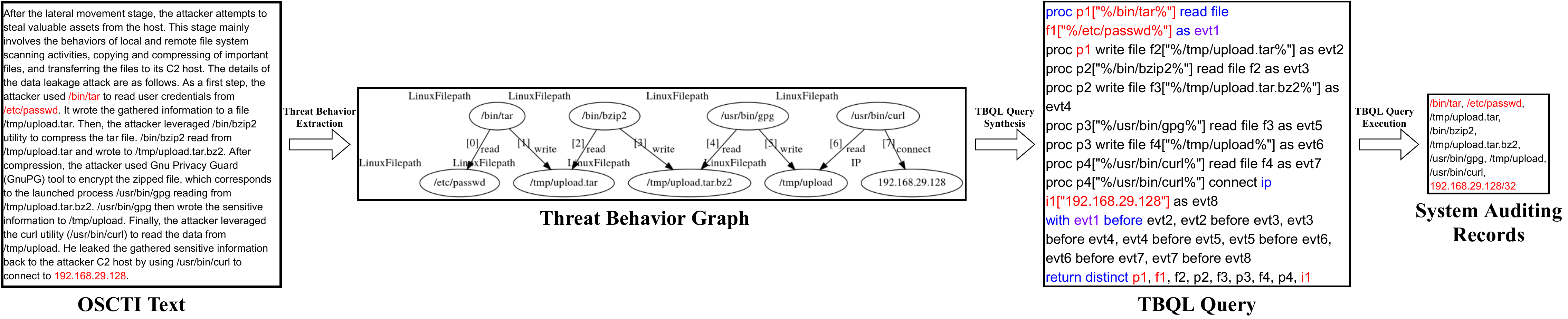}
    \caption{An example data leakage attack case demonstrating the whole processing pipeline of \tool}
    \label{fig:demo}
\end{figure*}

\cti~\cite{os-cti} is a form of evidence-based knowledge and has received growing attention from the community, enabling companies and organizations to gain visibility into the fast-evolving threat landscape.
Commonly, knowledge about threats is presented in a vast number of publicly available \cti sources.
Structured \cti feeds~\cite{phishtank,stix,misp,openioc} have primarily focused on Indicators of Compromise (IOCs)
\cite{liao2016acing}, which are forensic artifacts of an intrusion such as malicious file/process names, virus signatures, and IPs/domains of botnets.
Though useful in capturing fragmented views of threats, these disconnected IOCs lack the capability to uncover the complete threat scenario as to how the threat unfolds into multiple steps.
Consequently, 
defensive solutions that rely on these low-level, fragmented indicators~\cite{splunk-spl,elastic-siem} can be easily evaded when the attacker re-purposes the tools and changes their signatures.
In contrast, unstructured \cti reports~\cite{alienvault,securelist,krebsonsecurity} contain more comprehensive knowledge about threats.
For example, descriptive relationships between IOCs contain knowledge about \emph{multi-step threat behaviors} 
(\eg ``read'' relationship between two IOCs ``/bin/tar'' and ``/etc/passwd'' in \cref{fig:demo}), which is critical to uncovering the complete threat scenario.
Besides, such 
connected threat behaviors
are tied to the attacker's goals and thus more difficult to change.
Unfortunately, prior approaches do not provide an automated way to harvest such knowledge and use it for threat hunting.

\myparatight{Challenges}
In this work, we seek to design automated techniques to (1) extract knowledge about threat behaviors (IOCs and their relationships) from unstructured \cti reports, and (2) use the extracted knowledge to facilitate threat hunting.
We identify two major challenges.
First, accurately extracting threat knowledge from natural-language \cti text is not trivial.
This is due to the presence of massive nuances particular to the security context, such as special characters (\eg dots, underscores) in IOCs.
These nuances confuse most NLP modules (\eg sentence segmentation, tokenization), making existing
information extraction tools ineffective~\cite{angeli2015leveraging,openie5}.
Second, system auditing often produces a huge amount of daily logs (0.5 GB $\sim$ 1 GB for 1 enterprise host~\cite{reduction}), and hence threat hunting is a procedure of ``finding a needle in a haystack''. 
Such big data poses challenges for solutions to store and query the data efficiently
to hunt for malicious activities.
To meet the requirement of timely threat hunting, knowledge extraction from \cti text also needs to be efficient.

\begin{figure*}[t]
    \centering
    \includegraphics[width=0.63\linewidth]{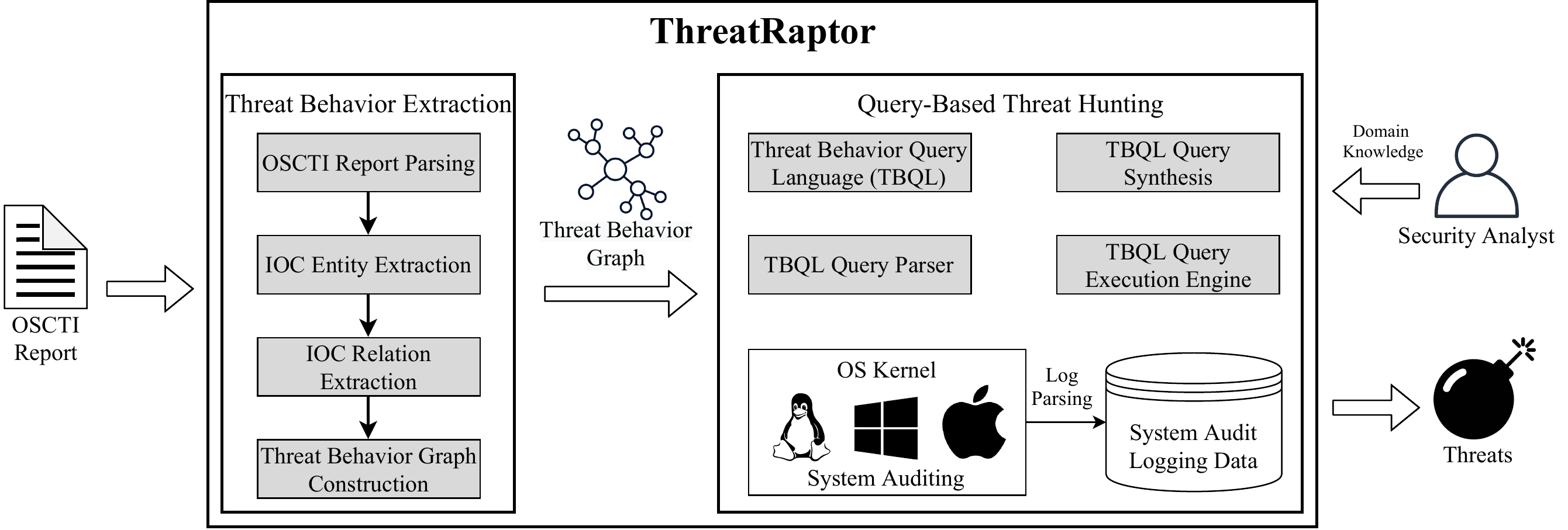}
    \caption{The architecture of \tool}
    \label{fig:arch}
\end{figure*}

\myparatight{Contribution}
We propose \tool, a system that facilitates threat hunting in computer systems using \cti.
We built \tool ($\sim25$K LOC) upon mature system auditing frameworks~\cite{auditd,etw,sysdig} 
for system audit logging data collection (\cref{subsec:auditing}), and databases
(PostgreSQL~\cite{postgresql}, Neo4j~\cite{neo4j})
for data storage (\cref{subsec:storage}). 
This enables our system to leverage the services provided by these mature infrastructures, such as data management, querying, and recovery.
Besides, \tool has three novel designs:

(1) \emph{Unsupervised, Light-Weight, and Accurate NLP Pipeline for Threat Behavior Extraction:} 
\tool employs a specialized NLP pipeline that targets the unique problem of IOC and IOC relation extraction from \cti text, which has not been studied in prior work.
To handle nuances and meet the requirement of timely threat hunting, the pipeline adopts a collection of techniques (\eg IOC protection, dependency parsing-based IOC relation extraction) to achieve accurate and efficient threat behavior extraction. 
The extracted threat behaviors are represented in a structured \emph{threat behavior graph}, in which nodes represent IOCs and edges represent IOC relations. 
Compared to the unstructured \cti text, such structured threat behavior representation is more amenable to automated processing and integration (\cref{subsec:extraction}).

(2) \emph{Domain-Specific Query Language \& Query Synthesis}:
To facilitate threat hunting over system audit logging data, \tool has an efficient query subsystem that employs a concise and expressive domain-specific query language, \emph{Threat Behavior Query Language (\lang)}, to query the 
log data stored in database backends.
\lang is a declarative language that 
integrates a collection of critical primitives for threat hunting in computer systems.
For example, \lang treats system entities (\ie files, processes, network connections) and system events (\ie file events, process events, network events) as first-class citizens, and provides explicit constructs for entity/event types, event operations, and event path patterns.
With \lang, complex multi-step system 
behaviors can be easily specified and searched (\cref{subsec:language}).
To bridge the threat behavior graph with the query sub-system, \tool employs a \emph{query synthesis mechanism} that automatically synthesizes a \lang query from the constructed graph.
This way, external knowledge about threat behaviors
can be automatically integrated in threat hunting.
No prior work has proposed a query language for threat hunting that supports the same set of features as supported in \lang, and has considered the automation of the threat hunting procedure via query synthesis (\cref{subsec:synthesis}).

It is important to note that \tool also supports human-in-the-loop analysis via query editing: the security analyst can further revise the synthesized query to encode domain knowledge about the specific enterprise.
In practice, threat hunting is an iterative process that involves multiple rounds of query editing and execution, and the conciseness and declarative nature of \lang make this process efficient.

(3) \emph{Efficient Query Execution:}
To query the big 
data efficiently, \tool employs specialized optimizations for data storage and query execution engine.
Specifically, \tool employs data reduction techniques to merge excessive system events 
while preserving adequate information.
To execute a \lang query, \tool decomposes it into parts and compiles each part into a semantically equivalent data query (\ie a small SQL~\cite{sql} or Cypher~\cite{cypher} query that will be executed in PostgreSQL or Neo4j databases).
\tool then employs a \emph{scheduling algorithm} to schedule the execution of these data queries,
based on their estimated pruning power and semantic dependencies.
Compared to the naive plan that compiles the \lang query into a giant SQL or Cypher query to execute, our execution plan avoids the weaving of many joins and constraints together (which often leads to slow performance) and leverages the query semantics to speed up the execution.
In addition to the exact search mode, \tool supports a \emph{fuzzy search mode} based on inexact graph pattern matching, by extending \cite{milajerdi2019poirot}.
This mode generalizes to cases where the searched graph pattern in a \lang query deviates from the ground truth (could due to IOC changes or structural differences), which improves the 
generality of \lang queries in threat hunting (\cref{subsec:execution}).

\myparatight{Evaluation}
We deployed \tool on a physical testbed and performed a broad set of attacks to evaluate 
the system.
The evaluation results demonstrate that:
(1) \tool is able to accurately extract threat behaviors from \cti text ($96.64\%$ F1 for IOC extraction, $92.59\%$ F1 for IOC relation extraction), performing much better than general information extraction approaches ($<5\%$ F1);
(2) \tool is able to accurately find malicious system activities using \cti text ($98.34\%$ F1);
(3) the entire pipeline of \tool is efficient.
The threat behavior extraction and query synthesis parts take $0.52$s on average.
For query execution (in exact search mode), \lang queries execute $22.7$x faster than SQL queries for PostgreSQL backend, and $9.1$x faster than Cypher queries for Neo4j backend;
(4) \lang queries are more concise than SQL queries ($>2.8$x) and Cypher queries ($>2.2$x).
To the best of our knowledge, \tool is \emph{the first system that bridges \cti with system auditing to facilitate cyber threat hunting in computer systems}.
A system demo video is available at \cite{threatraptor-demo}.

\section{System Overview}
\label{sec:overview}

\cref{fig:arch} shows the architecture of \tool, which consists of two subsystems: 
(1) a threat behavior extraction pipeline for automated threat knowledge extraction, and 
(2) a 
query subsystem built upon system auditing, which provides a domain-specific query language, \lang, to hunt for threats in computer systems.
In the query subsystem, monitoring agents built upon mature
frameworks~\cite{auditd,etw,sysdig}
are deployed across hosts to collect system audit logging data.
The collected data is sent to the central database for storage.
Given an input \cti report, \tool first extracts IOCs (\eg file names/paths, IPs) and their relations, and constructs a threat behavior graph. 
\tool then synthesizes a \lang query from the constructed graph, and executes the query to find the matched system auditing records.
The security analyst can optionally revise the synthesized query to encode 
domain knowledge.
In the situation where the \cti report is not available, \tool can be used as a proactive threat hunting tool with \lang queries manually constructed.

\myparatight{Demo Example}
\cref{fig:demo} shows an example data leakage attack case demonstrating the whole pipeline. 
The case was constructed based on the Cyber Kill Chain framework~\cite{killchain} and CVE~\cite{cve}, and used in our evaluation (\ie Case \emph{ra_2}).
As we can see, the threat behavior graph clearly shows how the threat unfolds into multiple connected steps, where each step is represented by an IOC node-edge triplet.
Furthermore, each edge is associated with a sequence number indicating the step order. 
Such sequential information is essential to uncovering the correct threat scenario and has not been considered in prior work~\cite{angeli2015leveraging,openie5}.
The synthesized \lang query further encodes the threat knowledge into formal query constructs, which is more amenable to human-in-the-loop analysis and iterative exploration.
Nodes and edges in the threat behavior graph are synthesized into system entities and system event patterns in the \lang query, and sequence numbers of edges are used to construct a {\tt with} clause that specifies the temporal order constraints of system event patterns.
By default, the synthesized \lang query specifies the default attributes of all system entities (\ie ``name'' for files, ``exename'' for processes, and ``dstip'' for network connections) in the {\tt return} clause.

\myparatight{Threat Model}
Our threat model follows the prior work on system auditing~\cite{backtracking,lee2013high,liu2018priotracker,gao2018aiql,gao2018saql,milajerdi2019poirot}.
We assume an attacker that attacks the computer system from outside: the attacker either utilizes the vulnerabilities in the system or convinces the user to download files with malicious payload.
We also assume that OS kernels and kernel-layer auditing frameworks
are part of our trusted computing base, and the system audit logging data collected from kernel space is not tampered.
We also do not consider attacks that do not go through kernel-layer auditing (\eg side channel attacks, memory-based attacks) and thus cannot be captured by system auditing frameworks.

\begin{table}[t]
\centering
\begin{adjustbox}{width=.77\linewidth}
    \begin{tabular}{ll}
        \hline
        \textbf{Event Category} &  \textbf{Relevant System Call} \\ \hline
        ProcessToFile & read, readv, write, writev, execve, rename \\ 
        ProcessToProcess & execve, fork, clone \\
        ProcessToNetwork & read, readv, recvfrom, recvmsg, sendto, write, writev \\ \hline
    \end{tabular}
\end{adjustbox}
\caption{Representative system calls processed}
\label{tab:syscalls}
\end{table}

\begin{table}[t]
\centering
\begin{adjustbox}{width=.77\linewidth}
    \begin{tabular}{ll}
      	\hline
			\textbf{Entity}     & \textbf{Attributes} \\ \hline
			File                & Name, Path, User, Group \\
			Process             & PID, Executable Name, User, Group, CMD \\ 
			Network Connection  & SRC/DST IP, SRC/DST Port, Protocol \\ \hline
    \end{tabular}
\end{adjustbox}
\caption{Representative attributes of system entities}
\label{tab:entity-attributes}
\end{table}

\begin{table}[!t]
	\centering
	\begin{adjustbox}{width=.77\linewidth}
		\begin{tabular}{l|l}
			\hline
            \textbf{Operation Type}      & Read, Write, Execute, Start, End, Rename \\
			\textbf{Time}		    & Start Time, End Time, Duration \\
			\textbf{Misc.}		    & Subject ID, Object ID, Data Amount, Failure Code \\ \hline
		\end{tabular}
	\end{adjustbox}
	\caption{Representative attributes of system events}
	\label{tab:event-attributes}
\end{table}

\section{Design of \tool}
\label{sec:approach}

\subsection{System Auditing}
\label{subsec:auditing}

\tool leverages mature system auditing frameworks~\cite{auditd,etw,sysdig} to collect system-level audit logs about system calls from the OS kernel.
The collected kernel audit logs consist of system 
events that describe the interactions among system entities, which are crucial for security analysis.
As shown in previous studies~\cite{backtracking,lee2013high,liu2018priotracker,gao2018aiql,gao2018saql,milajerdi2019poirot},
on mainstream operating systems,
system entities in most cases are files, processes, and network connections, and the monitored system calls 
are mapped to three major types of system events: file access, processes creation and destruction, and network access. 
Hence, in \tool, we consider \emph{system entities} as \emph{files}, \emph{processes}, and \emph{network connections}.
We consider a \emph{system event} as the interaction between two system entities represented as $\langle$subject_entity, operation, object_entity$\rangle$, which consists of the initiator of the interaction, the type of the interaction, and the target of the interaction.
Subjects are processes originating from software applications (\eg Chrome), and objects can be files, processes, and network connections.
We categorize system events into three types according to the types of their object entities: \emph{file events}, \emph{process events}, and \emph{network events}.

\tool parses the collected audit logs of system calls (\cref{tab:syscalls}) into a sequence of system events among system entities, and extracts a set of attributes that are crucial for security analysis (\cref{tab:entity-attributes,tab:event-attributes}).
To uniquely identify system entities,
for a process entity, \tool uses the process executable name and PID as its unique identifier.
For a file entity, \tool uses the absolute path as its unique identifier. 
For a network connection entity, 
\tool uses the 5-tuple $\langle$srcip, srcport, dstip, dstport, protocol$\rangle$ as 
its unique identifier~\cite{liu2018priotracker}. 
Failing to distinguish different entities will cause problems in relating events to entities.

\subsection{Data Storage}
\label{subsec:storage}

\tool stores the parsed system entities and system events in databases, so that the system audit logging data can be persisted.
Prior work has modeled system audit logging data as either relational tables~\cite{gao2018aiql} or provenance graphs~\cite{liu2018priotracker}.
Inspired by such designs, \tool adopts two types of database models for its storage component: relational model and graph model.
Relational databases come with mature
indexing mechanisms and are scalable to massive data, which are suitable for queries that involve many joins and constraints.
Graph databases represent data as nodes and edges, 
which are suitable for queries that involve graph pattern search.

Currently, \tool adopts PostgreSQL~\cite{postgresql} for its relational
storage and Neo4j~\cite{neo4j} for its graph 
storage.
For PostgreSQL, \tool stores system entities and system events in separate tables.
For Neo4j, \tool stores system entities as nodes and system events as edges.
Data is replicated across the two databases, which supports the execution of different types of queries (\cref{subsec:execution}) and improves data availability.
Indexes are created on key attributes (\eg file name, process executable name, source/destination IP) for both databases to speed up the search.

\myparatight{Data Reduction}
\tool further reduces the data size before storing it in databases, so that the search can be done more efficiently while the critical information about malicious behaviors is still preserved.
System audit logging data 
often has many excessive events between the same entity pair~\cite{reduction}.
The reason is that OS typically finishes a read/write task (\eg file read/write) by distributing the data proportionally to multiple system calls.
Inspired by a log reduction work
\cite{reduction},
\tool merges the events between two entities using the following criteria:
(1) two events $e_1(u_1, v_1)$ and $e_2(u_2, v_2)$ ($u_1$, $u_2$ are subjects and $v_1$, $v_2$ are objects; suppose $e_1$ occurs before $e_2$) will be merged if: $u_1 = u_2$ \&\& $v_1 = v_2$ \&\& $e_1.operationType = e_2.operationType$ \&\& $0 \leq e_2.startTime - e_1.endTime \leq threshold$;
(2) the attributes of the merged event $e_{m}$ are updated as: $e_{m}.startTime = e_1.startTime$, $e_{m}.endTime = e_2.endTime$, $e_{m}.dataAmount = e_1.dataAmount + e_2.dataAmount$. 
We experimented with different threshold 
and chose $1$ second, as it has reasonable reduction performance in merging system events for file manipulations, file transfers, and network communications, with no false events generated.

\subsection{Threat Behavior Extraction}
\label{subsec:extraction}

\begin{algorithm}[t]
\scriptsize

\SetAlgoLined
\SetKwData{Doc}{document}\SetKwData{Blk}{block}\SetKwData{RR}{replacementRecord}\SetKwData{Sent}{sentence}\SetKwData{Tree}{tree}\SetKwData{Trees}{trees}\SetKwData{Treess}{all_block_trees}\SetKwData{Bg}{graph}\SetKwData{Ent}{entities}\SetKwData{AEnt}{all_iocs}\SetKwData{Rel}{ioc_relations}\SetKwData{ARel}{all_ioc_relations}
\SetKwFunction{BSeg}{SegmentBlock}\SetKwFunction{NEP}{ProtectIoc}\SetKwFunction{SSeg}{SegmentSentence}\SetKwFunction{DP}{ParseDependency}\SetKwFunction{RNEP}{RemoveIocProtection}\SetKwFunction{TL}{AnnotateTree}\SetKwFunction{Coref}{ResolveCoref}\SetKwFunction{LBSP}{SimplifyTree}\SetKwFunction{BGG}{ConstructBehaviorGraph}\SetKwFunction{SRE}{ScanMergeIoc}\SetKwFunction{AE}{MergeSimilarEntities}\SetKwFunction{ME}{ExtractIocRelation}\SetKwFunction{GG}{ConstructGraph}
\SetKwInOut{Input}{Input}
\SetKwInOut{Output}{Output}
\BlankLine
\Input{OSCTI Text: \Doc}
\Output{Threat Behavior Graph: \Bg}
\BlankLine

Initialize \Treess\;
Initialize \ARel\;
\For{\Blk in \BSeg{\Doc}}{
    Initialize \Trees\;
    \Blk, \RR $\leftarrow$ \NEP{\Blk}\;
    \For{\Sent in \SSeg{\Blk}}{
        \Tree $\leftarrow$ \DP{\Sent}\;
        Align \RR with \Tree\;
        \Tree $\leftarrow$ \RNEP{\Tree, \RR}\;
        \Tree $\leftarrow$ \TL{\Tree}\;
        \Tree $\leftarrow$ \LBSP{\Tree}\;
        Add \Tree to \Trees\;
    }
    \For{\Tree in \Trees}{
        \Tree $\leftarrow$ \Coref{\Tree, \Trees}\;
    }
    Add all \Tree in \Trees to \Treess\; 
}

\AEnt $\leftarrow$ \SRE{\Treess}\;
\For{\Tree in \Trees}{
    \Rel $\leftarrow$ \ME{\Tree, \Trees, \AEnt}\;
    Add \Rel to \ARel\;
}
\Bg $\leftarrow$ \GG{\AEnt, \ARel}\;

\caption{Threat Behavior Extraction Pipeline}
\label{alg:threat-behavior-extraction}

\end{algorithm}

As mentioned in \cref{sec:intro}, massive nuances exist in \cti text (\eg dots, underscores in IOCs), which limit the performance of most NLP modules and existing information extraction tools~\cite{angeli2015leveraging,openie5}.
To address the unique challenge, \tool employs a specialized NLP pipeline to handle nuances and accurately extract IOCs and their relations to construct a threat behavior graph.
Furthermore, our pipeline is unsupervised and light-weight.
\cref{alg:threat-behavior-extraction} gives the pipeline:

\emph{Step 1: Block Segmentation (Line 3):}
We segment an article into blocks,
and extract IOCs and their relations from each block.
Later on, when we construct the threat behavior graph, we will link the same IOCs that appear across multiple blocks.

\emph{Step 2: IOC Recognition and IOC Protection (Line 5):}
We construct a set of regex rules by extending an open-source IOC parser~\cite{ioc-parser}
(we made improvements to improve its coverage, \eg distinguishing Linux/Windows file paths)
to recognize different types of IOCs (\eg file name, file path, IP).
Furthermore, we protect the security context by replacing the IOCs with a dummy word (we use the word ``something'') and leave a replacement record.
This makes the NLP modules designed for processing general text work well for \cti text.

\emph{Step 3: Sentence Segmentation (Line 6):}
We segment a block into sentences using a sentence segmentation component~\cite{spacy}.

\emph{Step 4: Dependency Parsing (Line 7):}
We construct a dependency tree for each sentence using a dependency parsing component~\cite{spacy} 
pretrained on a large general corpus.
We then use the replacement record of IOCs to restore the security context by replacing the dummy word with the original IOCs.

\emph{Step 5: Tree Annotation (Line 10):}
Among all nodes in the dependency trees, 
there are some nodes whose associated tokens are particularly useful for coreference resolution and \emph{relation extraction} (\eg IOCs, candidate IOC relation verbs, pronouns).
We annotate these nodes of interest in the trees.
We curated a list of keywords that correspond to candidate IOC relation verbs (\eg ``read'', ``write'', ``download'', ``open'').

\emph{Step 6: Tree Simplification (Line 11):}
We simplify the annotated trees by removing irrelevant nodes and paths (\ie removing the trees without any candidate IOC relation verbs or the paths without any IOC nodes).
This step does not influence the extraction outcome, but helps speed up the performance.

\emph{Step 7: Coreference Resolution (Line 14):}
Across all trees of all sentences within a block, we resolve the coreferenced nodes for the same IOC by checking their POS tags and dependencies,
and create connections between the nodes in the trees.
After this step, we have a set of final annotated, simplified dependency trees for the \cti text.

\emph{Step 8: IOC Scan and Merge (Line 16):}
As the same IOC may appear across different blocks in different phrases, we scan all IOCs in the dependency trees of all blocks, and merge similar ones based on both the character-level overlap and the semantic similarity of word vectors (we used word vectors in spaCy~\cite{spacy}).
This is different from Step 7, which performs coreference resolution within a block.
After this step, we have a set of final IOCs served as nodes in the threat behavior graph.

\emph{Step 9: IOC Relation Extraction (Line 18):}
We present the details of our \emph{dependency parsing-based IOC relation extraction algorithm}:
(1) For each dependency tree, we enumerate all pairs of IOCs nodes;
(2) Then, for each pair of IOC nodes, we check whether they satisfy the subject-object relation by considering their dependency types in the tree.
In particular, we consider three parts of their dependency path: one common path from the root to the LCA (Lowest Common Ancestor); two individual paths from the LCA to each of the nodes, and construct a set of dependency type rules to do the checking;
(3) Next, for the pair that passes the checking, we extract its relation verb by first scanning all the annotated candidate verbs (annotations are done in Step 5 using our curated list) in the aforementioned three parts of dependency path, and then selecting the one that is closest to the object IOC node;
(4) The candidate IOC node pair and the selected verb (after lemmatization) form the final IOC entity-relation triplet.
Note that for a token to be output as the final relation verb, it needs to be both covered by our keyword list and form the correct subject-verb-object relation with the IOC node pair tokens.

\emph{Step 10: Threat Behavior Graph Construction (Line 20):}
We iterate over all IOC entity-relation triplets sorted by the occurrence offset of the relation verb in \cti text, and construct a threat behavior graph.
Each edge in the graph is associated with a sequence number, indicating the step order.

\begin{BNF}[t]
\scriptsize

\begin{mdframed}
    \setlength{\grammarparsep}{0pt} 
    \setlength{\grammarindent}{8em}
    
    \begin{grammar}
    <tbql> ::= (<global_filter>)* (<patt>)+ <rel>? <return>
    
    <global_filter> ::= <attr_exp> | <wind>
    
    <patt> ::= <entity> (<op_exp> | <op_path>) <entity> <patt_id>? <wind>?
    
    <entity> ::= <entity_type> <id> (`[' <attr_exp>`]')?
    
    <entity_type> ::= `file' | `proc' | `ip'
    
    <op_exp> ::= <op>
    \alt `!'<op_exp>
    \alt <op_exp> (`&&' | `||') <op_exp>
    \alt `(' <op_exp> `)'
    
    <op> ::= `read' | `write' | `start' | `execute'|...
    
    <op_path> ::= (`~>' | `->') (`('<int>? `~'? <int>? `)')? (`[' <op_exp> `]')?
    
    <patt_id> ::= `as' <id> (`[' <attr_exp>`]')?
    
    <attr_exp> ::= <attr> <bop> <val>
    \alt `!'? <val> 
    \alt <attr> `not'? `in' <val_set>
    \alt <attr_exp> (`&&' | `||') <attr_exp>
    \alt `(' <attr_exp> `)'
    
    <attr> ::= <id> (`.' <id>)?
    
    <wind> ::= `from' <datetime> `to' <datetime>
    \alt (`at' | `before' |  `after') <datatime>
    \alt `last' <num> <time_unit>

    <rel> ::= `with' <id> (`before' | `after' | `within') (`[' <num> `-' <num> <time_unit>`]')? <id>
    \alt `with' <attr> <bop> <attr>
    
    <return> ::= `return' `distinct'? <attr> (`,' <attr>)*

    \end{grammar}

\end{mdframed}

\caption{Representative BNF grammar of \lang}\label{bnf:grammar}
\end{BNF}

\subsection{Threat Behavior Query Language (\lang)}
\label{subsec:language}

\tool provides a domain-specific language, \lang (\cref{bnf:grammar}), to facilitate threat hunting over system audit logging data.
Compared to low-level and verbose general query languages (SQL~\cite{sql}, Cypher~\cite{cypher}), 
\lang 
integrates a collection of critical primitives,
making it easy to specify complex multi-step system behaviors for hunting.

\emph{(1) Event Pattern Syntax:}
The basic syntax of \lang follows our prior work~\cite{gao2018aiql}, which specifies one or more system event patterns in the format of $\langle$subject_entity, operation, object_entity$\rangle$, 
with optional filters on temporal and attribute relationships between event patterns.
System entities have explicit types and identifiers, with optional filters on attributes.
Essentially, the specified event patterns form a subgraph of system events to be searched.
\cref{fig:demo} shows an example.

Specifically, in \cref{bnf:grammar}, the \emph{$\langle$patt$\rangle$} rule specifies an event pattern, including the subject/object entity (\emph{$\langle$entity$\rangle$}), the operation (\emph{$\langle$op_exp$\rangle$}), the pattern ID (\emph{$\langle$patt_id$\rangle$}), and the optional time window (\emph{$\langle$wind$\rangle$}). 
The \emph{$\langle$entity$\rangle$} rule specifies the entity type, the entity ID, and the optional attribute filter expression (\emph{$\langle$attr_exp$\rangle$}). 
Operators (\eg logical, comparison) are supported in \emph{$\langle$op_exp$\rangle$} and \emph{$\langle$attr_exp$\rangle$} to form complex expressions (\eg \incode{proc p[pid = 1 && exename = "\%chrome.exe\%"] read || write file f}, where \incode{\%} matches any character sequence).
The \emph{$\langle$wind$\rangle$} rule specifies a time window that narrows down the search.
The \emph{$\langle$global_filter$\rangle$} rule specifies the global filters for all event patterns.
The \emph{$\langle$rel$\rangle$} rule specifies the relationship between event patterns. 
\lang supports two types of relationships: temporal relationship (\eg \incode{evt1 before[0-5 min] evt2} specifies a temporal order of events), and attribute relationship (\eg \incode{p1.pid = p2.pid} specifies that the two processes have the same PID).
The \emph{$\langle$return$\rangle$} rule specifies the attributes of the matched events for return.

In addition, \lang provides different types of syntactic sugars to facilitate the query construction:

\begin{itemize}[noitemsep, topsep=1pt, partopsep=1pt, listparindent=\parindent, leftmargin=*]
    \item \emph{Default attributes for system entities:}
    default attribute names will be inferred if the user only specifies attribute values in an event pattern, or entity IDs in the {\tt return} clause. 
    We select the most commonly used attributes in security analysis as default attributes:
    ``name'' for files, ``exename'' for processes, and ``dstip'' for network connections.
    
    \item \emph{Entity ID reuse:} reusing an entity ID in multiple event patterns implicitly indicates that the entities are the same.
\end{itemize}

For example, in the \lang query in \cref{fig:demo}, \incode{proc p1["\%/bin/tar\%"]}, \incode{file f1["\%/etc/passwd\%"]}, \incode{ip i1["192.168.29.128"]}, and \incode{return p1} will be inferred as
\incode{proc p1[exename = "\%/bin/tar\%"]}, \incode{file f1[name = "\%/etc/passwd\%"]}, \incode{ip i1[dstip = "192.168.29.128"]}, and \incode{return p1.exename}.
Besides, the entity ID \incode{p1} is used in both \incode{evt1} and \incode{evt2}, indicating the same system entity.

\emph{(2) Variable-Length Event Path Pattern Syntax:}
In addition to the basic event pattern syntax, \tool uniquely provides an advanced syntax that specifies various types of variable-length paths of system event patterns.
The \emph{$\langle$op_path$\rangle$} rule gives the core syntax, which provides several alternatives:

\begin{itemize}[noitemsep, topsep=1pt, partopsep=1pt, listparindent=\parindent, leftmargin=*]
    \item \incode{proc p \~>[read] file f}: a path of arbitrary length from a process entity \incode{p} to a file entity \incode{f}. The operation type of the final hop (\ie system event where \incode{f} is an object) is \incode{read}.
    
    \item \incode{proc p \~>(2~4)[read] file f}: the path has a minimum length of $2$ and a maximum length of $4$.
    
    \item \incode{proc p \~>(2~)[read] file f}: the path has a minimum length of $2$. The maximum length is not restricted.
    
    \item \incode{proc p \~>(~4)[read] file f}: the path has a maximum length of $4$. The minimum length is $1$.

    \item \incode{proc p ->[read] file f}: the path has a length of $1$. This is semantically equivalent to the basic event pattern syntax, \eg \incode{proc p read file f}. 
    The difference lies in the execution: this length-1 event path pattern will be compiled into a Cypher data query executed on the Neo4j database, while the basic event pattern will be compiled into a SQL data query executed on the PostgreSQL database.

    \item \incode{proc p \~> file f}: the operation type of the final hop is omitted, indicating that the search allows any operation type.
\end{itemize}

This syntax is particularly useful when doing query synthesis: in certain cases, an edge in the threat behavior graph (hence a threat step between two IOCs in \cti text) may correspond to a path of system events in system audit logs.
This happens often when intermediate processes are created to chain system events, but are omitted in the \cti text by the human writer. 
With this syntax, the information flow between two system entities can be easily specified and the semantic gap between the \cti text and the system audit logs can be bridged.
We use ``\lang pattern'' to refer to both the event pattern and the variable-length event path pattern.

\subsection{\lang Query Synthesis}
\label{subsec:synthesis}

To facilitate threat hunting with \cti, \tool provides a query synthesis mechanism that automatically synthesizes a \lang query from the threat behavior graph.

\emph{Step 1: Pre-Synthesis Screening \& IOC Relation Mapping:}
One challenge in query synthesis is the semantic gap between the types of IOCs and IOC relations, and the types of system entities and their operations.
To bridge the gap, \tool first performs a pre-synthesis screening to filter out nodes (and connected edges) in the threat behavior graph whose associated IOC types are not currently captured by the system auditing component (\eg registry entries).
Then, for each remaining edge, \tool maps its associated IOC relation to the \lang operation type (\eg \emph{$\langle$op$\rangle$} rule in \cref{bnf:grammar}).
We constructed a set of rules for IOC relation mapping, which consider both the semantic meaning of the IOC relation and the types of the connected IOC nodes.
For example, the ``download'' relation between two ``Filepath'' IOCs will be mapped to the ``write'' operation in \lang, indicating a process writes data to a file.
In contrast, the ``download'' relation from a ``Filepath'' IOC to an ``IP'' IOC will be mapped to the ``read'' operation in \lang, indicating a process reads data from a network connection.
\tool further filters out edges whose associated IOC relations do not match any rules.

\emph{Step 2: \lang Pattern Synthesis:}
For each node in the threat behavior graph, \tool synthesizes a \lang system entity (\ie rule \emph{$\langle$entity$\rangle$}) and assigns a unique entity ID:
(1) for a source node, \tool synthesizes a process entity;
(2) for a sink node, \tool synthesizes a network connection entity if its associated IOC type is an IP. 
Otherwise, \tool synthesizes a file entity or a process entity depending on the associated IOC relation of the edge.
\tool then synthesizes the attribute of the entity using the associated IOC content.
Wildcard operators \incode{\%} are added around the attribute string by default.

\tool synthesize a \lang pattern (\ie rule \emph{$\langle$patt$\rangle$}) by connecting the synthesized \lang subject \& object entities and the mapped \lang operation.
By default, an event pattern is synthesized.
System administrator can configure the system to synthesize a variable-length event path pattern.

\emph{Step 3: \lang Pattern Relationship Synthesis:}
For \lang event patterns, 
\tool synthesizes their temporal relationships by following an ascending order of the sequence numbers of corresponding edges in the threat behavior graph. 
For variable-length event path patterns, this step is omitted since event paths in \lang do not have temporal relationships.

\emph{Step 4: \lang Return Synthesis:}
To synthesize the \lang return clause, \tool by default appends all entity IDs to the ``return'' string. 
Default attribute names will be inferred when the query is executed and the corresponding attribute values will be returned (\ie \lang syntactic sugars).

\cref{fig:demo} shows an example \lang query synthesized using the default synthesis plan.
In addition, \tool supports user-defined synthesis plans to overwrite the default plan and synthesize attributes that are supported but not captured in the threat behavior graph (\eg hostname, time window).

\subsection{\lang Query Execution}
\label{subsec:execution}

To efficiently execute a \lang query with many \lang patterns (could be a mix of event patterns and variable-length event path patterns), 
\tool extends our prior work~\cite{gao2018aiql} by (1) compiling each \lang pattern into a semantically equivalent SQL or Cypher data query, and (2) scheduling the execution of these data queries in different database backends (\ie PostgreSQL and Neo4j) based on their estimated pruning power and semantic dependencies (\cite{gao2018aiql} does not involve variable-length event path patterns and Neo4j backend).
Specifically, for an event pattern, \tool compiles it into a SQL data query, so that the mature indexing mechanism and the efficient support for joins in relational databases can be leveraged.
The compiled SQL query joins two system entity tables with one system event table, and applies the filters in the WHERE clause.
For a variable-length event path pattern, since it is difficult to perform graph pattern search using SQL, \tool compiles it into a Cypher data query by leveraging Cypher's path pattern syntax~\cite{cypher}.

We now present our \emph{data query scheduling algorithm}: For each \lang pattern, \tool computes a pruning score 
by counting the number of constraints declared; a \lang pattern with more constraints has a higher score.
For a variable-length event path pattern, we additionally consider the length of the path when computing the score; a pattern with a smaller maximum path length has a higher score.
Then, when scheduling the execution of the data queries, \tool considers both the pruning scores and the pattern dependencies: 
if two \lang patterns have dependencies (\eg connected by the same system entity), \tool will first execute the data query whose associated pattern has a higher pruning score, and then use the execution results to constrain the execution of the other data query (by adding filters).
This way, complex \lang queries with various \lang patterns
can be efficiently executed in different database backends seamlessly.

In addition to the 
search mode based on exact matching, \tool supports a \emph{fuzzy search mode} based on inexact graph pattern matching~\cite{milajerdi2019poirot}.
The user can use this mode as an alternative when the exact search mode fails to retrieve meaningful results, allowing 
the generality of searching while at the cost of a longer execution time.
Intuitively, a \lang query specifies a subgraph of system events to be searched,
and inexact graph pattern matching can be naturally leveraged in query execution to enable fuzzy search.
In the current design, our fuzzy search mode 
leverages Poirot~\cite{milajerdi2019poirot}
to search for both node-level alignment and graph-level alignment:
(1) For node-level alignment, we use Levenshtein distance~\cite{levenshtein1966binary} to perform similarity matching of IOC strings specified in the \lang query and attributes of system entities stored in the database, so that typos or small changes in IOCs can still retrieve the correct system entities;
(2) For graph-level alignment, we match the subgraph pattern specified in the \lang query with the provenance graph of system events.
We borrow Poirot's idea that measures the potential attacker influence by the number of compromised ancestor processes.
By calculating Poirot's graph alignment scores for all candidate alignments, \tool's query execution engine produces an exhaustive searching result for aligned subgraphs of system events, and returns the entity attributes specified in the {\tt return} clause as final results.
\cref{subsubsec:eval-rq4,sec:literature} include evaluation results and comparison with Poirot.

\section{Evaluation}
\label{sec:eval}

We built \tool ($\sim25$K LOC) upon several tools:
Sysdig~\cite{sysdig} 
for system auditing, 
PostgreSQL~\cite{postgresql}
and Neo4j~\cite{neo4j}
for system audit logging data storage,
Python and spaCy~\cite{spacy} for threat behavior extraction,
ANTLR 4~\cite{antlr} 
for \lang language parser, 
and Java for the whole system.

We deployed \tool on a physical testbed to collect real system audit logs and hunt for malicious activities.
We evaluated \tool on a broad set of attack cases.
In total, the audit logs used in our evaluations contain $47,688,033$ system entities
and $55,840,381$ system events. 
We aim to answer the following research questions:
\begin{itemize}[noitemsep, topsep=1pt, partopsep=1pt, listparindent=\parindent, leftmargin=*]
    \item \textbf{RQ1:} How accurate is \tool in extracting threat behaviors from \cti text compared to general information extraction approaches?
    
    \item \textbf{RQ2:} How accurate is \tool in finding malicious system activities using \cti text?
    
    \item \textbf{RQ3:} How efficient is \tool in extracting threat behaviors from \cti text, constructing a threat behavior graph, and synthesizing a \lang query?
    
    \item \textbf{RQ4:} How efficient is \tool in executing \lang queries over the big system audit logging data?

    \item \textbf{RQ5:} How concise is \lang in specifying malicious system behaviors compared to general-purpose query languages?
\end{itemize}

RQ1 aims evaluate the accuracy of \tool in threat behavior extraction.
RQ2 aims to evaluate the end-to-end accuracy of \tool in threat hunting using \cti.
RQ3 aims to evaluate the efficiency of \tool in threat behavior extraction, threat behavior graph construction, and \lang query synthesis.
RQ4 aims to evaluate the efficiency of \tool in \lang query execution, and measure the performance speedup achieved by the \lang query scheduler.
RQ5 aims to evaluate the conciseness of \lang in expressing complex system behaviors.

\subsection{Evaluation Setup}
\label{subsec:eval-setup}

The deployed server has an Intel(R) Xeon(R) CPU E5-2637 v4 (3.50GHz), 256GB RAM running 64bit Ubuntu 18.04.1.
The server is frequently used by $>15$ active users to perform various daily tasks, including file manipulation, text editing, and software development.
To evaluate \tool, we constructed an evaluation benchmark of $18$ attack cases from two sources: 
$15$ cases released in the DARPA TC dataset~\cite{darpatc}, and $3$ multi-step intrusive attacks that we performed ourselves on the testbed based on the Cyber Kill Chain framework~\cite{killchain} and CVE~\cite{cve}.
When we perform the attacks and conduct the evaluations, the sever continues to serve other users.
This setup ensures that enough noise of benign background traffic is collected in together with malicious activities, representing the real-world deployment.
Furthermore, benign activities significantly outnumber attack activities ($55$ million vs. thousands), demonstrating the challenge in threat hunting. 
\cref{tab:case-name} shows the list of cases.
The total monitoring length is $41$ days for DARPA TC cases and $16$ hours for our three attacks.

\begin{table}[t]
\begin{adjustbox}{width=.9\linewidth,center}
\begin{tabular}{ll}
\hline

\textbf{Case ID}         & \textbf{Case Name}  \\\hline
tc\_clearscope\_1     & 20180406 1500 ClearScope – Phishing E-mail Link   \\
tc\_clearscope\_2     & 20180411 1400 ClearScope – Firefox Backdoor w/ Drakon In-Memory    \\
tc\_clearscope\_3     & 20180413 ClearScope                                                 \\
tc\_fivedirections\_1 & 20180409 1500 FiveDirections – Phishing E-mail w/ Excel Macro       \\
tc\_fivedirections\_2 & 20180411 1000 FiveDirections – Firefox Backdoor w/ Drakon In-Memory \\
tc\_fivedirections\_3 & 20180412 1100 FiveDirections – Browser Extension w/ Drakon Dropper  \\
tc\_theia\_1          & 20180410 1400 THEIA – Firefox Backdoor w/ Drakon In-Memory      \\
tc\_theia\_2          & 20180410 1300 THEIA - Phishing Email w/ Link                        \\
tc\_theia\_3          & 20180412 THEIA - Browser Extension w/ Drakon Dropper             \\
tc\_theia\_4          & 20180413 1400 THEIA - Phishing E-mail w/ Executable Attachment    \\
tc\_trace\_1          & 20180410 1000 TRACE – Firefox Backdoor w/ Drakon In-Memory        \\
tc\_trace\_2          & 20180410 1200 TRACE – Phishing E-mail Link                          \\
tc\_trace\_3          & 20180412 1300 TRACE – Browser Extension w/ Drakon Dropper         \\
tc\_trace\_4          & 20180413 1200 TRACE – Pine Backdoor w/ Drakon Dropper             \\
tc\_trace\_5          & 20180413 1400 TRACE – Phishing E-mail w/ Executable Attachment    \\
password\_crack                 & Password Cracking After Shellshock Penetration                                               \\
data\_leak                 & Data Leakage After Shellshock Penetration                                                \\
vpnfilter                 & VPNFilter     
        \\                                         
\hline
\end{tabular}
\end{adjustbox}
\caption{18 attack cases in our evaluation benchmark}
\label{tab:case-name}
\end{table}

\subsubsection{DARPA TC Attack Cases}
\label{subsec:tc}

We selected 15 cases from the DARPA TC Engagement 3 data release~\cite{darpatc},
which cover various combinations of OSs (\eg Linux, Windows, Android), vulnerabilities (\eg Nginx backdoor, Firefox backdoor, browser extension), and exploits (\eg Drakon APT, micro APT, phishing email with malicious Excel attachment).

Specifically, the dataset consists of the captured audit logs of six performer systems (ClearScope, FiveDirections, THEIA, TRACE, CADETS, and TA5.2) under the penetration of the red team using different attack strategies, which include both benign and malicious system activities.
The dataset also includes a ground-truth report that has attack descriptions for the cases.
After examining the descriptions and the logs, we found that 
the logs for TA5.2 are missing in the released dataset and the logs for CADETS lack key attributes (\eg file name).
This makes us unable to confirm the attack ground truth to conduct faithful evaluations.
Thus, we do not consider these cases in our evaluations.
Nevertheless, similar attacks were already performed for other performer systems and their descriptions and logs are covered in our evaluations.
For the other four performer systems, we selected all the $15$ attack cases in our evaluation benchmark.
For each case, we parsed the provided audit logs and loaded the data in \tool's databases.
We then extracted the attack description text from the ground-truth report and use it as input to \tool.

\begin{table*}[t]
\begin{adjustbox}{width=.8\textwidth,center}
\begin{tabular}{lcccccc}
\hline

\textbf{Approaches}               
& \textbf{Entity Precision} 
& \textbf{Entity Recall}
& \textbf{Entity F1}
& \textbf{Relation Precision}
& \textbf{Relation Recall}
& \textbf{Relation F1} \\
\hline
\tool                             & 96.00\%	& 97.30\%	& \textbf{96.64\%}	& 96.15\%	& 89.29\%	& \textbf{92.59\%}
\\
\tool - IOC Protection            & 94.12\%	& 43.24\%	& 59.26\%	& 100.00\%	& 8.93\%	& 16.39\%                                                          \\
Stanford Open IE                 & 1.82\%	& 14.86\%	& 3.24\%	& 0.00\%	& 0.00\%	& 0.00\%                \\
Stanford Open IE + IOC Protection & 4.39\%	& 36.49\%	& 7.84\%	& 0.88\%	& 8.93\%	& 1.59\%                                                     \\
Open IE 5                         & 0.25\%	& 1.35\%	& 0.43\%	& 0.00\%	& 0.00\%	& 0.00\%                                                      \\
Open IE 5 + IOC Protection       & 3.49\%	& 20.27\%	& 5.95\%	& 0.00\%	& 0.00\%	& 0.00\%                                 \\
\hline
\end{tabular}
\end{adjustbox}
\caption{Precision, recall, and F1 of IOC entity extraction and IOC relation extraction of \tool and baseline approaches. The results are aggregated overall all 18 cases. 
}
\label{tab:nlp-accuracy-aggregate}
\end{table*}

\subsubsection{Multi-Step Intrusive Attack Cases}
\label{subsubsec:ra}

To increase the coverage of our benchmark, we constructed $3$ multi-step intrusive attack cases, based on 
CVE~\cite{cve} and capture the important traits of attacks depicted in the Cyber Kill Chain framework~\cite{killchain} (\eg including the stages of initial penetration, data exfiltration).
We performed these attacks on the testbed and collected system audit logs.
The attack description texts were constructed according to the way the attacks were performed.

\myparatight{Attack 1: Password Cracking After Shellshock Penetration}
The attacker penetrates into the victim host (\ie the testbed) by exploiting the Shellshock vulnerability~\cite{shellshock}. 
After penetration, the attacker first connects to cloud services (Dropbox) and downloads an image where C2 (Command \& Control) server's IP address is encoded in the EXIF metadata. 
This behavior is a common practice shared by APT attacks~\cite{vpnfilter}
to evade 
DNS blacklisting based detection systems.
Using the IP, the attacker downloads a password cracker from the C2 server to the victim host, and then runs the password cracker against password shadow files to extract clear text.

\myparatight{Attack 2: Data Leakage After Shellshock Penetration}
After the reconnaissance, 
the attacker attempts to steal all the valuable assets from the victim host. 
This stage mainly involves the behaviors of local and remote file system scanning activities, copying and compressing of important files, and transferring the files to the C2 server.
The attacker scans the file system, scrapes files into a single compressed file, and transfers it back to the C2 server.

\myparatight{Attack 3: VPNFilter}
The attacker seeks to maintain a direct connection to the victim host from the C2 server. 
The attacker utilizes the notorious VPNFilter malware~\cite{vpnfilterschenier} which infected millions of IoT devices by exploiting a number of known or zero-day vulnerabilities.
After the initial penetration on the victim host, the attacker downloads the VPNFilter stage 1 malware from the C2 server, which accesses a public image repository to get an image.
In the EXIF metadata of the image, the IP address for the stage 2 server is encoded.
The stage 1 malware then downloads the VPNFilter stage 2 malware from the stage 2 server, and executes it to launch the VPNFilter attack, which establishes a direct connection to the C2 server.

\subsection{Evaluation Results}
\label{subsec:eval-results}

\subsubsection{RQ1: Accuracy of Threat Behavior Extraction}
\label{subsubsec:eval-rq1}

To evaluate the accuracy of \tool in extracting threat behaviors from \cti text, we labeled the \cti texts based on the ground truth and measure the precision, recall, and F1 
of the extracted IOC entities and IOC relations.
We compare \tool with two state-of-the-art open information extraction
approaches for entity and relation extraction from general text:
Stanford Open IE~\cite{angeli2015leveraging} and Open IE 5~\cite{openie5}.
Furthermore, we are interested in studying the effect of IOC Protection on the accuracy of IOC entity and relation extraction. 
Thus, we also compare \tool with the version of \tool without IOC Protection, Stanford Open IE with IOC Protection, and Open IE 5 with IOC Protection. 

\cref{tab:nlp-accuracy-aggregate} shows the precision, recall, and F1 score aggregated over all evaluation cases.
We have the following observations:
(1) \tool achieves the highest precision, recall, and F1 score for both IOC entity extraction and IOC relation extraction. In particular, \tool has $96.64\%$ F1 for IOC entity extraction and $92.59\%$ F1 for IOC relation extraction. 
In contrast, the scores of Stanford Open IE and Open IE 5 are very low: $< 5\%$ F1 for IOC entitiy extraction and $0\%$ F1 for IOC relation extraction.
These results demonstrate the effectiveness of \tool's specialized threat behavior extraction NLP pipeline in processing \cti text; 
(2) When removing IOC Protection, the scores of \tool drop significantly ($59.26\%$ F1 for IOC entity extraction and $16.39\%$ F1 for IOC relation extraction). 
The reason for the accuracy drop is that if the \cti text is processed directly by an NLP component without first applying IOC Protection, the sentence segmentation component and the tokenizer will break the IOC entities (\eg file paths, process executable names, IPs) into pieces, making it impossible to annotate the pieces and analyzing the correct grammatical structure of sentences.
This demonstrates the effectiveness of IOC Protection in protecting the security context in \cti text;
(3) When adding IOC Protection, the scores of Stanford Open IE and Open IE 5 increase a bit, but not much. 
This again demonstrates the effectiveness of IOC Protection in improving the accuracy of other NLP components.
Though, as these approaches target general information extraction instead of threat behavior extraction from \cti text, their performance is limited.

\subsubsection{RQ2: Accuracy of Threat Hunting}
\label{subsubsec:eval-rq2}

\begin{table}[t]
\begin{adjustbox}{width=0.7\linewidth,center}
\begin{tabular}{lcc}
\hline

\multirow{2}{*}{\textbf{Case}}         &  \textbf{Precision}         & \textbf{Recall} \\
& \textbf{TP/(TP+FP)} &  \textbf{TP/(TP+FN)} \\
\hline
tc\_clearscope\_1     & 6/6  & 6/6    \\
tc\_clearscope\_2     & 3/3  & 3/3    \\
tc\_clearscope\_3     & 1/1  & 1/1    \\
tc\_fivedirections\_1 & 51/51  & 51/51    \\
tc\_fivedirections\_2 & 3/3  & 3/3    \\
tc\_fivedirections\_3 & 0/0  & 0/3    \\
tc\_theia\_1          & 3/3  & 3/3    \\
tc\_theia\_2          & 115/115  & 115/115    \\
tc\_theia\_3          & 3/3  & 3/3    \\
tc\_theia\_4          & 421/421  & 421/421    \\
tc\_trace\_1          & 39/39  & 39/76    \\
tc\_trace\_2          & 7/7  & 7/7    \\
tc\_trace\_3          & 0/0  & 0/2    \\
tc\_trace\_4          & 1/1  & 1/3    \\
tc\_trace\_5          & 578/578  & 578/578    \\
password\_crack       & 10/10  & 10/12    \\
data\_leak            & 6/6  & 6/8    \\
vpnfilter             & 178/178  & 178/178    \\         
\textbf{Total}        & 1425/1425 = 100.00\%   & 1425/1473 = 96.74\% \\   
\hline
\end{tabular}
\end{adjustbox}
\caption{Precision and recall of \tool in finding malicious system events}
\label{tab:end2end-accuracy}
\end{table}

\begin{table*}[]
\begin{adjustbox}{width=.9\textwidth,center}
\begin{tabular}{lcccccccc}
\hline

& \multicolumn{3}{c}{\textbf{\tool}}
& \multicolumn{1}{c}{\textbf{\tool $-$ IOC Protection}}
& \multicolumn{1}{c}{\textbf{Stanford Open IE}}
& \multicolumn{1}{c}{\textbf{Stanford Open IE $+$ IOC Protection}}
& \multicolumn{1}{c}{\textbf{Open IE 5}}
& \multicolumn{1}{c}{\textbf{Open IE 5 $+$ IOC Protection}}
\\\cline{2-9}

\multirow{-2}{*}{\textbf{Case}}
& \textbf{Text -> E. \& R.}
& \textbf{E. \& R. -> Graph}
& \textbf{Graph -> \lang}
& \textbf{Text -> E. \& R.}
& \textbf{Text -> E. \& R.}
& \textbf{Text -> E. \& R.}
& \textbf{Text -> E. \& R.} 
& \textbf{Text -> E. \& R.}
\\ \hline
tc\_clearscope\_1     & 0.43                                                     & 0.08                                                      & 0.00                                                                  & 0.42                                                     & 0.74                                                     & 0.74                                                     & 7.69                                                     & 7.69                                                     \\
tc\_clearscope\_2     & 0.42                                                     & 0.08                                                      & 0.00                                                                  & 0.39                                                     & 0.46                                                     & 0.46                                                     & 1.53                                                     & 1.53                                                     \\
tc\_clearscope\_3     & 0.37                                                     & 0.02                                                      & 0.00                                                                  & 0.24                                                     & 0.76                                                     & 0.76                                                     & 15.29                                                    & 15.29                                                    \\
tc\_fivedirections\_1 & 0.33                                                     & 0.08                                                      & 0.01                                                                  & 0.35                                                     & 0.61                                                     & 0.61                                                     & 7.84                                                     & 7.84                                                     \\
tc\_fivedirections\_2 & 0.39                                                     & 0.08                                                      & 0.01                                                                  & 0.38                                                     & 0.56                                                     & 0.56                                                     & 0.67                                                     & 0.67                                                     \\
tc\_fivedirections\_3 & 0.36                                                     & 0.08                                                      & 0.00                                                                  & 0.35                                                     & 0.63                                                     & 0.63                                                     & 0.50                                                     & 0.50                                                     \\
tc\_theia\_1          & 0.52                                                     & 0.09                                                      & 0.01                                                                  & 0.46                                                     & 0.59                                                     & 0.59                                                     & 16.53                                                    & 16.53                                                    \\
tc\_theia\_2          & 0.52                                                     & 0.08                                                      & 0.01                                                                  & 0.47                                                     & 0.31                                                     & 0.31                                                     & 48.85                                                    & 48.85                                                    \\
tc\_theia\_3          & 0.56                                                     & 0.09                                                      & 0.01                                                                  & 0.51                                                     & 0.72                                                     & 0.72                                                     & 7.79                                                     & 7.79                                                     \\
tc\_theia\_4          & 0.26                                                     & 0.09                                                      & 0.01                                                                  & 0.27                                                     & 0.73                                                     & 0.73                                                     & 0.19                                                     & 0.19                                                     \\
tc\_trace\_1          & 0.44                                                     & 0.11                                                      & 0.01                                                                  & 0.48                                                     & 1.06                                                     & 1.06                                                     & 4.01                                                     & 4.01                                                     \\
tc\_trace\_2          & 0.45                                                     & 0.16                                                      & 0.01                                                                  & 0.44                                                     & 0.63                                                     & 0.63                                                     & 47.13                                                    & 47.13                                                    \\
tc\_trace\_3          & 0.31                                                     & 0.09                                                      & 0.00                                                                  & 0.31                                                     & 1.10                                                     & 1.10                                                     & 1.00                                                     & 1.00                                                     \\
tc\_trace\_4          & 0.46                                                     & 0.09                                                      & 0.00                                                                  & 0.46                                                     & 0.96                                                     & 0.96                                                     & 1.76                                                     & 1.76                                                     \\
tc\_trace\_5          & 0.42                                                     & 0.08                                                      & 0.00                                                                  & 0.43                                                     & 0.91                                                     & 0.91                                                     & 41.92                                                    & 41.92                                                    \\
password_crack                 & 0.43                                                     & 0.09                                                      & 0.01                                                                  & 0.44                                                     & 0.87                                                     & 0.87                                                     & 25.54                                                    & 25.54                                                    \\
data_leak                 & 0.47                                                     & 0.09                                                      & 0.01                                                                  & 0.51                                                     & 0.84                                                     & 0.84                                                     & 10.06                                                    & 10.07                                                    \\
vpnfilter                 & 0.35                                                     & 0.08                                                      & 0.01                                                                  & 0.30                                                     & 1.15                                                     & 1.16                                                     & 4.05                                                     & 4.05                                                     \\
\textbf{Total}        & 7.50                                                     & 1.55                                                      & 0.11                                                                  & 7.21                                                     & 13.63                                                    & 13.64                                                    & 242.35                                                   & 242.36                                                   \\
\textbf{Average}      & 0.42                                                     & 0.09                                                      & 0.01                                                                  & 0.40                                                     & 0.76                                                     & 0.76                                                     & 13.46                                                    & 13.46    \\
\hline
\end{tabular}
\end{adjustbox}
\caption{Execution time (second) of different stages of \tool: threat behavior extraction (text -> E. \& R.), threat behavior graph construction (E. \& R. -> graph), and \lang query synthesis (graph -> \lang)}
\label{tab:nlp-perf}
\end{table*}

To measure the end-to-end accuracy of \tool in threat hunting, for each attack case, we compare the system events found by the event patterns in the synthesized \lang query, 
and the ground-truth system events that are related to the attack.
\cref{tab:end2end-accuracy} shows the precision and recall.
We have the following observations:
(1) \tool is able to accurately find malicious system events using \cti texts,
achieving $100\%$ precision, $96.74\%$ recall, and 
$98.34\%$
F1.
This is largely due to the high accuracy achieved by \tool's threat behavior extraction pipeline;
(2) Though some excessive event patterns may be occasionally synthesized (\eg in \emph{password\_crack}, one excessive event pattern is synthesized: \incode{proc p3["\%/tmp/libfoo.so\%"] write file f2["\%/tmp/john.zip\%"] as evt5}), the design of \tool ensures that these excessive event patterns will rarely retrieve benign activities.
The reason is because these excessive patterns have IOCs as subject/object constraints, which are extracted by a set of highly-precise regex rules in \tool.
As a result, very few benign activities are falsely retrieved (\eg $0$ false positive rate in our evaluation benchmark);
(3) For queries that have false negatives, the primary reason is due to the semantic ambiguity in query synthesis for certain IOC relations. 
For example, in \emph{tc\_trace\_1}, there is an edge pointing from the ``Filepath'' IOC ``/home/admin/cache'' to itself with the ``run'' relation.
Both the IOC and the relation are corrected extracted from \cti text.
However, when performing query synthesis, there is no way to differentiate whether it represents a file event \incode{proc p1["\%/home/admin/cache\%"] execute file f1["\%/home/admin/cache\%"]} or a process event \incode{proc p1["\%/home/admin/cache\%"] start proc p2["\%/home/admin/cache\%"]}, as both events are related to process creation.
The default synthesis plan in \tool synthesizes the first pattern, while for this case, the second pattern has matched ground-truth system events.
As a result, $37$ system events are missed.
One way to mitigate this is to let the security analyst revise the query to improve the coverage, and the synthesized event patterns
serve as a base for exploration.

It is worth mentioning that the three cases for ClearScope were conducted on Android OS and the ground-truth system events have Android package names as process executable names (\eg \incode{proc p1["\%com.android.defcontainer\%"] open file f1["\%MsgApp-instr.apk\%"]}), which are different from other cases in which process executables are normal Linux/Windows files. 
Nevertheless, \tool is able to accurately extract such information 
and use the information to find the malicious system events, thanks to the coverage of a wide range of IOC types and IOC relations in \tool's threat behavior extraction pipeline.

\subsubsection{RQ3: Efficiency of Threat Behavior Extraction}
\label{subsubsec:eval-rq3}

\cref{tab:nlp-perf} shows the execution time
of different stages of \tool: threat behavior extraction, threat behavior graph construction, and query synthesis. 
For threat behavior extraction, we also compare with other baseline 
approaches.
We have the following observations:
(1) \tool is efficient in processing the input \cti texts, constructing threat behavior graphs, and synthesizing \lang queries. The average time for the three stages is $0.52$s;
(2) Stanford Open IE and Open IE 5 are more expensive in extracting threat behaviors compared to \tool ($0.76$s and $13.46$s vs. $0.42$s), since these general information extraction approaches spend a long time analyzing texts that are unrelated to threat behaviors;
(3) IOC Protection adds trivial overhead.

\subsubsection{RQ4: Efficiency of \lang Query Execution}
\label{subsubsec:eval-rq4}

\begin{table}[t]
\begin{adjustbox}{width=\linewidth,center}
\begin{tabular}{lcccccccc}
\hline

& \multicolumn{2}{c}{\textbf{\lang}}
& \multicolumn{2}{c}{\textbf{SQL}}
& \multicolumn{2}{c}{\textbf{\lang (length-1 path)}}
& \multicolumn{2}{c}{\textbf{Cypher}}
\\\cline{2-9}

\multirow{-2}{*}{\textbf{Case}}         
& \textbf{20-r mean}         
& \textbf{20-r std}
& \textbf{20-r mean}         
& \textbf{20-r std}
& \textbf{20-r mean}         
& \textbf{20-r std}
& \textbf{20-r mean}         
& \textbf{20-r std}   \\
\hline
tc\_clearscope\_1     & 1.07                              & 0.14                             & 1.41                              & 0.54                             & 3.86                              & 0.21                             & 3.91                              & 0.19                             \\
tc\_clearscope\_2     & 1.39                              & 0.16                             & 1.34                              & 0.12                             & 4.14                              & 0.33                             & 3.93                              & 0.20                             \\
tc\_clearscope\_3     & 0.92                              & 0.15                             & 0.90                              & 0.11                             & 3.51                              & 0.24                             & 3.47                              & 0.17                             \\
tc\_fivedirections\_1 & 2.48                              & 0.04                             & 32.24                             & 0.48                             & 5.38                              & 0.38                             & 44.79                             & 1.01                             \\
tc\_fivedirections\_2 & 1.79                              & 0.16                             & 1.94                              & 0.12                             & 4.52                              & 0.31                             & 4.50                              & 0.36                             \\
tc\_fivedirections\_3 & 1.46                              & 0.14                             & 1.87                              & 0.17                             & 3.89                              & 0.28                             & 5.40                              & 0.36                             \\
tc\_theia\_1          & 3.86                              & 0.08                             & 43.15                             & 0.55                             & 10.41                             & 0.38                             & 234.31                            & 5.31                             \\
tc\_theia\_2          & 1.91                              & 0.17                             & 1.88                              & 0.17                             & 5.17                              & 0.41                             & 77.66                             & 1.11                             \\
tc\_theia\_3          & 4.43                              & 0.40                             & 12.07                             & 0.34                             & 9.84                              & 0.35                             & 32.66                             & 0.59                             \\
tc\_theia\_4          & 4.37                              & 0.15                             & 5.28                              & 0.29                             & 8.54                              & 0.32                             & 8.09                              & 0.43                             \\
tc\_trace\_1          & 44.21                             & 0.57                             & 85.63                             & 1.91                             & 82.16                             & 0.81                             & 366.11                            & 18.74                            \\
tc\_trace\_2          & 50.66                             & 0.77                             & 52.94                             & 0.42                             & 85.90                             & 0.66                             & 348.23                            & 1.94                             \\
tc\_trace\_3          & 1.97                              & 0.05                             & 1.95                              & 0.06                             & 2.25                              & 0.01                             & 2.24                              & 0.01                             \\
tc\_trace\_4          & 37.66                             & 0.37                             & 38.88                             & 0.26                             & 89.39                             & 0.85                             & 91.33                             & 0.53                             \\
tc\_trace\_5          & 5.43                              & 0.10                             & 13.79                             & 0.37                             & 6.46                              & 0.12                             & 6.19                              & 0.11                             \\
password_crack                 & 1.52                              & 0.20                             & 40.32                             & 0.45                             & 6.06                              & 0.34                             & 57.70                             & 1.02                             \\
data_leak                 & 1.45                              & 0.43                             & 3,456.12                          & 67.43                            & 6.28                              & 0.67                             & 1,803.14                          & 34.92                            \\
vpnfilter                 & 1.60                              & 0.29                             & 18.44                             & 0.28                             & 5.28                              & 0.42                             & 11.04                             & 0.61                             \\
\textbf{Total}        & 168.18                            & \multicolumn{1}{l}{}             & 3,810.17                          & \multicolumn{1}{l}{}             & 343.04                            & \multicolumn{1}{l}{}             & 3,104.68                          & \multicolumn{1}{l}{}             \\
\textbf{Average}      & 9.34                              & \multicolumn{1}{l}{}             & 211.68                            & \multicolumn{1}{l}{}             & 19.06                             & \multicolumn{1}{l}{}             & 172.48                            & \multicolumn{1}{l}{}            \\
\hline
\end{tabular}
\end{adjustbox}
\caption{Execution time (second) of queries in \lang, SQL, \lang in length-1 event path pattern syntax, and Cypher. Each query was executed for 20 rounds.}
\label{tab:query-perf}
\end{table}

We measure the runtime performance of \tool in executing \lang queries, particularly the performance speedup provided by the \lang query scheduler in different database backends.
To prepare for evaluation, for each case, we construct four types of semantically equivalent queries according to the corresponding synthesized \lang query by \tool:

\begin{enumerate}[noitemsep, topsep=1pt, partopsep=1pt, listparindent=\parindent, label=(\alph*), leftmargin=*]

    \item \lang query using the event pattern syntax (\eg \incode{proc p open file f as evt}).

    \item SQL query that encodes all event patterns and filters in the FROM and WHERE clauses.

    \item \lang query using the length-1 event path pattern syntax (\eg \incode{proc p ->[open] file f as evt}).

    \item Cypher query that encodes all length-1 event path patterns and filters in the MATCH and WHERE clauses.
\end{enumerate}

All these four types queries
search for the same system behaviors and return the same results.
The difference lies in the query scheduler and the database:
Queries (a) and (b) are executed in PostgreSQL, and Queries (c) and (d) are executed in Neo4j.
Queries (a) and (c) benefit from optimizations in \lang query scheduler, and Queries (b) and (d) do not.

\cref{tab:query-perf} shows the execution time
of the 
queries aggregated over $20$ rounds.
We have the following observations:
(1) \lang query scheduler provided by \tool is generally more efficient than the query schedulers provided by PostgreSQL and Neo4j. 
Specifically, for PostgreSQL backend, \tool is $3810.17/168.18 = 22.7$x faster; for Neo4j backend, \tool is $3104.68/343.04 = 9.1$x faster;
(2) There also exist a few cases in which \lang queries run slightly slower than SQL queries and Cypher queries. 
Particularly, when the \lang query only contains $1$ pattern (\ie \emph{tc\_clearscope\_3}, \emph{tc\_trace\_3}), \lang query runs slower than SQL query and Cypher query as additional time is taken to parse the \lang query and compile into SQL or Cypher data queries.
When the number of patterns becomes large, SQL queries and Cypher queries become much slower (\eg \emph{data\_leak}), as these giant queries have many joins and constraints mixed together, which may suffer from indeterministic optimizations and take long to finish execution;
(3) PostgreSQL is generally faster than Neo4j, as relational databases have 
better support for joins;
(4) The standard deviation 
is small compared to the mean.
This indicates that the $20$-round mean values are representative.
These results demonstrates the superiority of \lang query scheduler in speeding up the execution of \lang queries in different database backends.

\begin{table}[t]
\begin{adjustbox}{width=\linewidth,center}

\begin{tabular}{lcccccccc}
\hline

& \multicolumn{4}{c}{\textbf{\tool-Fuzzy}}
& \multicolumn{4}{c}{\textbf{Poirot}}
\\\cline{2-9}

\multirow{-2}{*}{\textbf{Case}}         
& \textbf{Loading}
& \textbf{Preprocessing}
& \textbf{Searching}
& \textbf{Total}
& \textbf{Loading}
& \textbf{Preprocessing}
& \textbf{Searching}
& \textbf{Total} \\
\hline

tc\_clearscope\_1     & 7.71                                          & 6.03                                                & 577.01                                       & 590.75                                      & 7.94                               & 6.08                                     & 542.59                                 & 556.61                                 \\
tc\_clearscope\_2     & 7.69                                          & 6.32                                                & \multicolumn{1}{l}{\textgreater{}3600}       & \multicolumn{1}{l}{\textgreater{}3600}      & 7.74                               & 6.20                                     & 431.22                                 & 445.17                                 \\
tc\_clearscope\_3     & 7.69                                          & 6.02                                                & 21.30                                        & 35.01                                       & 7.75                               & 5.86                                     & 22.21                                  & 35.81                                  \\
tc\_fivedirections\_1 & 18.11                                         & 12.99                                               & 58.33                                        & 89.42                                       & 18.15                              & 12.80                                    & 57.83                                  & 88.78                                  \\
tc\_fivedirections\_2 & 18.05                                         & 12.52                                               & 705.13                                       & 735.69                                      & 18.15                              & 12.36                                    & 663.13                                 & 693.64                                 \\
tc\_fivedirections\_3 & 18.09                                         & 12.37                                               & 33.28                                        & 63.74                                       & 18.08                              & 12.08                                    & 32.64                                  & 62.80                                  \\
tc\_theia\_1          & 37.25                                         & 34.10                                               & \multicolumn{1}{l}{\textgreater{}3600}       & \multicolumn{1}{l}{\textgreater{}3600}      & 37.35                              & 33.25                                    & \multicolumn{1}{l}{\textgreater{}3600} & \multicolumn{1}{l}{\textgreater{}3600} \\
tc\_theia\_2          & 37.36                                         & 33.06                                               & \multicolumn{1}{l}{\textgreater{}3600}       & \multicolumn{1}{l}{\textgreater{}3600}      & 37.36                              & 31.68                                    & \multicolumn{1}{l}{\textgreater{}3600} & \multicolumn{1}{l}{\textgreater{}3600} \\
tc\_theia\_3          & 36.98                                         & 35.15                                               & \multicolumn{1}{l}{\textgreater{}3600}       & \multicolumn{1}{l}{\textgreater{}3600}      & 37.34                              & 34.23                                    & \multicolumn{1}{l}{\textgreater{}3600} & \multicolumn{1}{l}{\textgreater{}3600} \\
tc\_theia\_4          & 37.02                                         & 33.10                                               & \multicolumn{1}{l}{\textgreater{}3600}       & \multicolumn{1}{l}{\textgreater{}3600}      & 37.31                              & 32.65                                    & \multicolumn{1}{l}{\textgreater{}3600} & \multicolumn{1}{l}{\textgreater{}3600} \\
tc\_trace\_1          & 297.35                                        & 510.99                                              & 249.71                                       & 1,058.05                                    & 296.65                             & 489.12                                   & 109.20                                 & 894.97                                 \\
tc\_trace\_2          & 295.01                                        & 504.08                                              & \multicolumn{1}{l}{\textgreater{}3600}       & \multicolumn{1}{l}{\textgreater{}3600}      & 297.02                             & 483.31                                   & \multicolumn{1}{l}{\textgreater{}3600} & \multicolumn{1}{l}{\textgreater{}3600} \\
tc\_trace\_3          & 295.30                                        & 340.92                                              & 111.55                                       & 747.77                                      & 297.22                             & 328.44                                   & 114.32                                 & 739.97                                 \\
tc\_trace\_4          & 294.23                                        & 412.82                                              & 301.30                                       & 1,008.34                                    & 297.43                             & 401.04                                   & 304.09                                 & 1,002.56                               \\
tc\_trace\_5          & 294.53                                        & 401.28                                              & \multicolumn{1}{l}{\textgreater{}3600}       & \multicolumn{1}{l}{\textgreater{}3600}      & 297.24                             & 389.22                                   & \multicolumn{1}{l}{\textgreater{}3600} & \multicolumn{1}{l}{\textgreater{}3600} \\
password\_crack       & 1.36                                          & 1.12                                                & 37.97                                        & 40.45                                       & 1.37                               & 1.54                                     & 22.02                                  & 24.93                                  \\
data\_leak            & 2.37                                          & 1.75                                                & 19.01                                        & 23.13                                       & 2.36                               & 1.72                                     & 18.46                                  & 22.55                                  \\
vpnfilter             & 2.07                                          & 1.83                                                & \multicolumn{1}{l}{\textgreater{}3600}       & \multicolumn{1}{l}{\textgreater{}3600}      & 2.07                               & 1.83                                     & 2,069.69                               & 2,073.59            \\

\hline
\end{tabular}
\end{adjustbox}

\caption{Execution time (second) of \tool's fuzzy search mode and Poirot~\cite{milajerdi2019poirot}.
}
\label{tab:fuzzy-perf}

\end{table}

\emph{Performance of \tool's Fuzzy Search Mode:}
We further study the performance of \tool's fuzzy search mode and Poirot~\cite{milajerdi2019poirot} (the difference is that Poirot does not search for all aligned system provenance subgraphs).
\cref{tab:fuzzy-perf} shows the execution time (second).
The execution consists of three parts: loading all system entities and system events from database into memory (loading time), constructing the provenance graph from system entities and system events (preprocessing time), and searching for alignments in the provenance graph (searching time).

We have the following observations:
(1) \tool's fuzzy search mode (\tool-Fuzzy) based on Poirot improves generality at the cost of efficiency.
In particular, \tool's exact search mode is a lot faster than \tool-Fuzzy and Poirot.
Besides, given that \tool-Fuzzy additionally performs an exhaustive search, it is within our expectation that \tool-Fuzzy in general runs longer than Poirot;
(2) There are several cases that both Poirot and \tool-Fuzzy cannot finish within 1 hour (\eg tc\_theia cases).
After profiling the execution,
we identify two major bottlenecks in Poirot's searching iterations that affect both approaches:
(a) a large number of candidate node alignments often result in a longer running time; 
(b) graph traversals required by candidate selection on dense graphs are time-consuming. 
For example, although the provenance graph of tc\_theia\_1 ($1.5$M nodes, $8.7$M edges) is a lot smaller than tc\_trace\_1 ($44.7$M nodes, $39.8$M edges), tc\_theia\_1 ($78$K alignments) has more candidate node alignments than tc\_trace\_1 ($131$ alignments). 
Furthermore, the average degree of tc\_theia is $5.9$ while the average degree of tc_trace is only $0.9$.
As a result, searching for aligned subgraphs for tc\_theia cases is more time-consuming;
(3) For cases that \tool-Fuzzy finishes within 1 hour, all ground-truth attack activities are found.

Based on the evaluation results, we recommend the user to use \tool's exact search mode if possible, which is much more efficient. 
The fuzzy search mode can be used as an alternative when the exact search mode fails to retrieve meaningful results.
How to enable more efficient inexact graph pattern matching (and exhaustive graph alignment search) in dense, high-alignment system provenance graphs is an open question, which we leave for future work.

\subsubsection{RQ5: Conciseness of \lang}
\label{subsubsec:eval-rq5}

\begin{table}[t]
\begin{adjustbox}{width=.92\linewidth,center}
\begin{tabular}{lccccccccc}
\hline
&
& \multicolumn{2}{c}{\textbf{\lang}}
& \multicolumn{2}{c}{\textbf{SQL}}
& \multicolumn{2}{c}{\textbf{\lang (length-l path)}}
& \multicolumn{2}{c}{\textbf{Cypher}} \\\cline{3-10}

\multirow{-2}{*}{\textbf{Case}}         
& \multirow{-2}{*}{\textbf{\# Patterns}}         
& \textbf{\# Chars}          
& \textbf{\# Words}        
& \textbf{\# Chars}          
& \textbf{\# Words}      
& \textbf{\# Chars}          
& \textbf{\# Words}      
& \textbf{\# Chars}          
& \textbf{\# Words} \\
\hline
tc\_clearscope\_1     & 2                                              & 131                                   & 30                                    & 584                                   & 102                                   & 145                                   & 32                                    & 530                                   & 84                                    \\
tc\_clearscope\_2     & 2                                              & 213                                   & 36                                    & 588                                   & 102                                   & 224                                   & 37                                    & 556                                   & 83                                    \\
tc\_clearscope\_3     & 1                                              & 93                                    & 16                                    & 218                                   & 40                                    & 97                                    & 16                                    & 205                                   & 30                                    \\
tc\_fivedirections\_1 & 3                                              & 219                                   & 47                                    & 737                                   & 134                                   & 234                                   & 48                                    & 674                                   & 105                                   \\
tc\_fivedirections\_2 & 3                                              & 164                                   & 39                                    & 750                                   & 133                                   & 182                                   & 41                                    & 665                                   & 105                                   \\
tc\_fivedirections\_3 & 2                                              & 111                                   & 25                                    & 376                                   & 70                                    & 125                                   & 26                                    & 330                                   & 51                                    \\
tc\_theia\_1          & 3                                              & 238                                   & 51                                    & 711                                   & 130                                   & 253                                   & 52                                    & 648                                   & 101                                   \\
tc\_theia\_2          & 3                                              & 235                                   & 51                                    & 888                                   & 152                                   & 253                                   & 53                                    & 803                                   & 124                                   \\
tc\_theia\_3          & 5                                              & 329                                   & 77                                    & 1385                                  & 252                                   & 361                                   & 81                                    & 1261                                  & 206                                   \\
tc\_theia\_4          & 2                                              & 195                                   & 39                                    & 562                                   & 102                                   & 206                                   & 40                                    & 530                                   & 83                                    \\
tc\_trace\_1          & 4                                              & 249                                   & 55                                    & 927                                   & 169                                   & 268                                   & 56                                    & 833                                   & 130                                   \\
tc\_trace\_2          & 3                                              & 264                                   & 57                                    & 933                                   & 168                                   & 282                                   & 59                                    & 882                                   & 140                                   \\
tc\_trace\_3          & 1                                              & 109                                   & 19                                    & 240                                   & 42                                    & 116                                   & 20                                    & 222                                   & 33                                    \\
tc\_trace\_4          & 3                                              & 187                                   & 40                                    & 656                                   & 120                                   & 202                                   & 41                                    & 599                                   & 91                                    \\
tc\_trace\_5          & 2                                              & 211                                   & 40                                    & 585                                   & 103                                   & 222                                   & 41                                    & 553                                   & 84                                    \\
password_crack                 & 8                                              & 551                                   & 122                                   & 1866                                  & 311                                   & 586                                   & 123                                   & 1639                                  & 252                                   \\
data_leak                 & 8                                              & 561                                   & 121                                   & 1841                                  & 328                                   & 596                                   & 122                                   & 1614                                  & 249                                   \\
vpnfilter                 & 5                                              & 400                                   & 80                                    & 1160                                  & 212                                   & 420                                   & 80                                    & 1057                                  & 162                                   \\
\textbf{Total}        & 60                                             & 4460                                  & 945                                   & 15007                                 & 2670                                  & 4772                                  & 968                                   & 13601                                 & 2113                                  \\
\textbf{Average}      & 3.33                                    & 247.78                       & 52.50                                & 833.72                           & 148.33                           & 265.11                           & 53.78                           & 755.61                           & 117.39\\
\hline
\end{tabular}
\end{adjustbox}
\caption{Conciseness of queries in \lang, SQL, \lang in length-1 event path pattern syntax, and Cypher}
\label{tab:conciseness}
\end{table}

For the four types of queries mentioned in RQ4, we further compare their conciseness by measuring the number of characters (excluding spaces and comments) and words.
\cref{tab:conciseness} shows the results.
We observe that:
(1) \lang is more concise than SQL and Cypher for all cases. Specifically, for \# characters, \lang is $15007/4460=3.4$x more concise than SQL and $13601/4772=2.9$x more concise than Cypher; for \# words, \lang is $2670/945=2.8$x more concise than SQL and $2113/968=2.2$x more concise than Cypher.
This is because \lang directly models the high-level, domain-specific concepts like system entities and system events, instead of the low-level concepts like tables or nodes/relationships;
(2) The conciseness saving of \lang compared to SQL and Cypher increases when more patterns are declared (\eg \emph{password\_crack}, \emph{data\_leak});
(3) Cypher queries are generally more concise than SQL queries. This is within our expectation as Cypher has a concise syntax to specify linked nodes and relationships, 
while SQL models everything as tables and has to explicitly specify table joins to represent system events.

\section{Discussion}
\label{sec:discussion}

\myparatight{Limitations}
As mentioned in \cref{sec:overview}, attacks on OS kernels, system auditing frameworks, and databases, and attacks that are not captured by system auditing
are 
not considered by \tool.
Besides, \tool's threat behavior extraction pipeline is not applicable
if the \cti text for the attack is not available or contains little useful information (\eg no IOCs, no sentence structures that contain IOC relations).
When there are deviations between the \cti text and the ground truth (\eg typos or changes in IOCs), \tool's exact search mode may miss attack activities.
In such cases, \tool's fuzzy search mode can be used as an alternative to increase the generality of searching.
Once some attack activities are found, the user can switch back to the exact search mode and revise the query (\eg adding event patterns) to expand the search for connected activities.

\myparatight{CTI Collection}
In general, CTI reports can be collected from various public sources (\aka \cti), such as security websites~\cite{alienvault,securelist} and blogs~\cite{krebsonsecurity}.
Enterprises may also have access to proprietary sources such as internal reports provided by domain experts, which might better reflect the particular enterprise environment.
This work does not target CTI collection.
Instead, \tool targets automated extraction of threat knowledge from an input report and use of the extracted knowledge for threat hunting.

\myparatight{Design Alternatives}
\tool currently leverages regex rules to extract IOCs and dependency parsing to extract IOC relations.
Besides IOCs, other types of entities may also exist in \cti text that constitute threat behaviors, such as threat actors (\eg CozyDuke~\cite{apt29}) and security tools (\eg Mimikatz~\cite{mimikatz}), which are hard to extract using fixed regex rules.
To extend the support for these entities and their relations, one approach is to adopt learning-based approaches to perform Named Entity Recognition (NER)~\cite{lample2016neural} and Relation Extraction (RE)~\cite{lin2016neural}. Different from \tool's current unsupervised NLP pipeline, these approaches are typically supervised, which require large annotated corpora for model training.
Such large annotated corpora is very costly to obtain manually.
In future work, we plan to explore techniques to programmatically synthesize annotations for \cti texts (\eg via data programming~\cite{ratner2016data}) and leverage learning-based approaches to expand our threat behavior extraction scope.

In query synthesis, \tool has a pre-synthesis screening step to filter out nodes in the threat behavior graph whose associated IOC types are not captured by the system auditing component.
In future work, we plan to expand our monitoring scope by including more types of entities and events (\eg Windows registry entries and Linux pipes).

\section{Related Work}
\label{sec:literature}

\myparatight{Forensic Analysis via System Audit Logs}
Research has proposed to leverage system audit logs for forensic analysis.
Causality analysis 
plays a critical role in identifying root causes and ramifications of attacks~\cite{backtracking}.
Efforts have been made to mitigate the dependency explosion problem by performing fine-grained causality analysis~\cite{lee2013high},
prioritizing dependencies~\cite{liu2018priotracker},
and reducing data size~\cite{reduction}. 
Besides, research has proposed to query system audit logs for attack investigation and anomaly detection~\cite{gao2018aiql,gao2018saql,pasquier2018runtime,gao2019query,gao2020querying}.
The scope of \tool is different from these works, as none of these works proposed to facilitate threat hunting via automated extraction of threat knowledge from \cti text and automated synthesis of threat hunting queries from the extracted knowledge.
Besides, the \lang query language provided in \tool has a set of features particularly designed for threat hunting (\eg variable-length event path pattern syntax) that are not supported in prior query tools.

Poirot~\cite{milajerdi2019poirot} is an approach for threat hunting that finds the aligned system provenance subgraph of an input query graph. 
Its core contribution is an inexact graph pattern matching algorithm for finding the alignment.
It is important to note that Poirot's scope is significantly different from \tool's scope:
Poirot does not search for all aligned subgraphs.
Instead, Poirot stops its searching iteration after finding the first acceptable alignment that surpasses a threshold.
This is different from the goal of \tool's query subsystem.
Besides, unlike \tool's automated threat behavior graph construction, Poirot's query graph requires non-trivial efforts of cyber analysts to manually construct it.
Furthermore, unlike \tool, Poirot does not involve database storage for storing the massive log data, a query language for proactive threat hunting, and a query synthesis mechanism for automating the process.
Nevertheless, Poirot's inexact graph pattern matching algorithm can be leveraged to improve the generality of searching:
\tool's current fuzzy search mode extends it to support exhaustive search.

\myparatight{\cti Analysis and Management}
Research progress has been made for automated \cti analysis, including extracting IOCs~\cite{liao2016acing},
extracting threat action terms from semi-structured Symantec reports~\cite{husari2017ttpdrill}, 
and measuring information inconsistency~\cite{dong2019towards}.
There also exist platforms and standards for \cti management and exchange~\cite{gao2021security,threatminer,misp,stix,openioc}.
\tool distinguishes from these works in the sense that it seeks to extract both IOCs and IOC relations from \cti text, 
and use the extracted knowledge for threat hunting.

\myparatight{Open Information Extraction}
Information extraction (IE) extracts structured information from unstructured natural language text. 
Open information extraction (Open IE)
is a new paradigm of IE that is not limited to a restricted set of target relations known in advance, but rather extracts all types of relations found in the text.
Research has proposed to leverage rule-based approaches or learning-based approaches for more accurate Open IE
~\cite{angeli2015leveraging,openie5}.
\tool distinguishes from these works in the sense that it focuses on threat behavior extraction from \cti text, which requires special designs to handle massive nuances particular to the security domain.

\section{Conclusion}
We have proposed \tool, a system that facilitates cyber threat hunting in computer systems using \cti.

\myparatight{Acknowledgement}
Peng Gao, Xiaoyuan Liu, and Dawn Song are supported in part by DARPA N66001-15-C-4066 and the 
Center for Long-Term Cybersecurity.
Fei Shao and Xusheng Xiao are supported in part by NSF CNS-2028748.
Zheng Qin and Fengyuan Xu are supported in part by NSFC-61872180, Jiangsu "Shuang-Chuang" Program, and Jiangsu "Six-Talent-Peaks" Program.
Prateek Mittal is supported in part by NSF CNS-1553437 and CNS-1704105, the ARL’s Army Artificial Intelligence Innovation Institute (A2I2), the Office of Naval Research Young Investigator Award, and the Army Research Office Young Investigator Prize.
Sanjeev R. Kulkarni is supported in part by the Center for Science of Information (CSoI), an NSF Science and Technology Center, under grant agreement CCF-0939370.

\bibliographystyle{IEEEtran}
\bibliography{refs}

\begin{thebibliography}{10}
\providecommand{\url}[1]{#1}
\csname url@samestyle\endcsname
\providecommand{\newblock}{\relax}
\providecommand{\bibinfo}[2]{#2}
\providecommand{\BIBentrySTDinterwordspacing}{\spaceskip=0pt\relax}
\providecommand{\BIBentryALTinterwordstretchfactor}{4}
\providecommand{\BIBentryALTinterwordspacing}{\spaceskip=\fontdimen2\font plus
\BIBentryALTinterwordstretchfactor\fontdimen3\font minus
  \fontdimen4\font\relax}
\providecommand{\BIBforeignlanguage}[2]{{%
\expandafter\ifx\csname l@#1\endcsname\relax
\typeout{** WARNING: IEEEtran.bst: No hyphenation pattern has been}%
\typeout{** loaded for the language `#1'. Using the pattern for}%
\typeout{** the default language instead.}%
\else
\language=\csname l@#1\endcsname
\fi
#2}}
\providecommand{\BIBdecl}{\relax}
\BIBdecl

\bibitem{target}
``{Target Data Breach Incident},''
  \url{http://www.nytimes.com/2014/02/27/business/target-reports-on-fourth-quarter-earnings.html?\_r=1}.

\bibitem{equifax}
``{The Equifax Data Breach},'' https://www.ftc.gov/equifax-data-breach.

\bibitem{backtracking}
S.~T. King and P.~M. Chen, ``Backtracking intrusions,'' in \emph{SOSP}, 2003.

\bibitem{lee2013high}
K.~H. Lee, X.~Zhang, and D.~Xu, ``High accuracy attack provenance via
  binary-based execution partition.'' in \emph{NDSS}, 2013.

\bibitem{liu2018priotracker}
Y.~Liu, M.~Zhang, D.~Li, K.~Jee, Z.~Li, Z.~Wu, J.~Rhee, and P.~Mittal,
  ``Towards a timely causality analysis for enterprise security,'' in
  \emph{NDSS}, 2018.

\bibitem{milajerdi2019poirot}
S.~M. Milajerdi, B.~Eshete, R.~Gjomemo, and V.~Venkatakrishnan, ``Poirot:
  Aligning attack behavior with kernel audit records for cyber threat
  hunting,'' in \emph{CCS}, 2019.

\bibitem{gao2018aiql}
P.~Gao, X.~Xiao, Z.~Li, F.~Xu, S.~R. Kulkarni, and P.~Mittal, ``{AIQL}:
  Enabling efficient attack investigation from system monitoring data,'' in
  \emph{{USENIX} ATC}, 2018.

\bibitem{gao2018saql}
P.~Gao, X.~Xiao, D.~Li, Z.~Li, K.~Jee, Z.~Wu, C.~H. Kim, S.~R. Kulkarni, and
  P.~Mittal, ``{SAQL}: A stream-based query system for real-time abnormal
  system behavior detection,'' in \emph{{USENIX} Security}, 2018.

\bibitem{pasquier2018runtime}
T.~Pasquier, X.~Han, T.~Moyer, A.~Bates, O.~Hermant, D.~Eyers, J.~Bacon, and
  M.~Seltzer, ``{Runtime Analysis of Whole-system Provenance},'' in \emph{CCS},
  2018.

\bibitem{splunk-spl}
``{Splunk Search Processing Language},''
  \url{https://www.splunk.com/en_us/resources/search-processing-language.html}.

\bibitem{elastic-siem}
``{Elastic SIEM},'' https://www.elastic.co/siem.

\bibitem{os-cti}
``{Open Source Threat Intelligence Feeds},''
  \url{https://www.senki.org/operators-security-toolkit/open-source-threat-intelligence-feeds/}.

\bibitem{phishtank}
``{PhishTank},'' \url{https://www.phishtank.com/}.

\bibitem{stix}
``{Structured Threat Information eXpression},'' http://stixproject.github.io/.

\bibitem{misp}
``{MISP - Open Source Threat Intelligence Platform \& Open Standards For Threat
  Information Sharing},'' \url{https://www.misp-project.org/}.

\bibitem{openioc}
``{The History of OpenIOC},''
  https://www.fireeye.com/blog/threat-research/2013/09/history-openioc.html.

\bibitem{liao2016acing}
X.~Liao, K.~Yuan, X.~Wang, Z.~Li, L.~Xing, and R.~Beyah, ``Acing the ioc game:
  Toward automatic discovery and analysis of open-source cyber threat
  intelligence,'' in \emph{CCS}, 2016.

\bibitem{alienvault}
``{AlienVault},'' \url{https://www.alienvault.com/blogs/labs-research/}.

\bibitem{securelist}
``{SecureList},'' \url{https://securelist.com/}.

\bibitem{krebsonsecurity}
``{KrebsonSecurity},'' \url{https://krebsonsecurity.com/}.

\bibitem{angeli2015leveraging}
G.~Angeli, M.~J.~J. Premkumar, and C.~D. Manning, ``Leveraging linguistic
  structure for open domain information extraction,'' in \emph{ACL}, 2015.

\bibitem{openie5}
``{Open IE 5},'' \url{https://github.com/dair-iitd/OpenIE-standalone}.

\bibitem{reduction}
Z.~Xu, Z.~Wu, Z.~Li, K.~Jee, J.~Rhee, X.~Xiao, F.~Xu, H.~Wang, and G.~Jiang,
  ``High fidelity data reduction for big data security dependency analyses,''
  in \emph{CCS}, 2016.

\bibitem{auditd}
``{The Linux Audit Framework},'' \url{https://github.com/linux-audit/}.

\bibitem{etw}
``{Event Tracing for Windows},''
  \url{https://docs.microsoft.com/en-us/windows/win32/etw/event-tracing-portal}.

\bibitem{sysdig}
``{Sysdig},'' \url{http://www.sysdig.org/}.

\bibitem{postgresql}
``{PostgreSQL},'' \url{http://www.postgresql.org/}.

\bibitem{neo4j}
``{Neo4j},'' \url{http://neo4j.com/}.

\bibitem{sql}
``{SQL: Structured Query Language},''
  \url{http://www.iso.org/iso/catalogue_detail.htm?csnumber=45498}.

\bibitem{cypher}
``{Cypher Query Language},'' \url{http://neo4j.com/developer/cypher/}.

\bibitem{threatraptor-demo}
``{ThreatRaptor Demo Video},'' \url{https://youtu.be/SrcTDQwRF_M}.

\bibitem{killchain}
``{Cyber Kill Chain},''
  https://www.lockheedmartin.com/en-us/capabilities/cyber/cyber-kill-chain.html.

\bibitem{cve}
``{Common Vulnerabilities and Exposures},'' \url{https://cve.mitre.org/}.

\bibitem{ioc-parser}
``{ioc-parser},'' \url{https://github.com/armbues/ioc_parser}.

\bibitem{spacy}
``{spaCy},'' \url{https://spacy.io/usage/linguistic-features}.

\bibitem{levenshtein1966binary}
V.~I. Levenshtein, ``Binary codes capable of correcting deletions, insertions,
  and reversals,'' in \emph{Soviet physics doklady}, 1966.

\bibitem{antlr}
``{ANTLR},'' \url{http://www.antlr.org/}.

\bibitem{darpatc}
``Transparent computing engagement 3 data release,''
  \url{https://github.com/darpa-i2o/Transparent-Computing/blob/master/README-E3.md}.

\bibitem{shellshock}
``{CVE}-2014-6271,''
  https://cve.mitre.org/cgi-bin/cvename.cgi?name=CVE-2014-6271.

\bibitem{vpnfilter}
``{VPNFilter: New Router Malware with Destructive Capabilities},''
  \url{https://symantec-enterprise-blogs.security.com/blogs/threat-intelligence/vpnfilter-iot-malware}.

\bibitem{vpnfilterschenier}
``{Router Vulnerability and the VPNFilter Botnet},''
  \url{https://www.schneier.com/blog/archives/2018/06/router_vulnerab.html}.

\bibitem{apt29}
``{APT29},'' \url{https://attack.mitre.org/groups/G0016/}.

\bibitem{mimikatz}
``{Mimikatz},'' \url{https://attack.mitre.org/software/S0002/}.

\bibitem{lample2016neural}
G.~Lample, M.~Ballesteros, S.~Subramanian, K.~Kawakami, and C.~Dyer, ``Neural
  architectures for named entity recognition,'' in \emph{NAACL-HLT}, 2016.

\bibitem{lin2016neural}
Y.~Lin, S.~Shen, Z.~Liu, H.~Luan, and M.~Sun, ``Neural relation extraction with
  selective attention over instances,'' in \emph{ACL}, 2016.

\bibitem{ratner2016data}
A.~J. Ratner, C.~M. De~Sa, S.~Wu, D.~Selsam, and C.~R\'{e}, ``Data programming:
  Creating large training sets, quickly,'' in \emph{NeurIPS}, 2016.

\bibitem{gao2019query}
P.~Gao, X.~Xiao, Z.~Li, K.~Jee, F.~Xu, S.~R. Kulkarni, and P.~Mittal, ``A query
  system for efficiently investigating complex attack behaviors for enterprise
  security,'' in \emph{{VLDB}}, 2019.

\bibitem{gao2020querying}
P.~Gao, X.~Xiao, D.~Li, K.~Jee, H.~Chen, S.~R. Kulkarni, and P.~Mittal,
  ``Querying streaming system monitoring data for enterprise system anomaly
  detection,'' in \emph{{ICDE}}, 2020.

\bibitem{husari2017ttpdrill}
G.~Husari, E.~Al-Shaer, M.~Ahmed, B.~Chu, and X.~Niu, ``Ttpdrill: Automatic and
  accurate extraction of threat actions from unstructured text of cti
  sources,'' in \emph{ACSAC}, 2017.

\bibitem{dong2019towards}
Y.~Dong, W.~Guo, Y.~Chen, X.~Xing, Y.~Zhang, and G.~Wang, ``Towards the
  detection of inconsistencies in public security vulnerability reports,'' in
  \emph{USENIX Security}, 2019.

\bibitem{gao2021security}
P.~Gao, X.~Liu, E.~Choi, B.~Soman, C.~Mishra, K.~Farris, and D.~Song, ``A
  system for automated threat intelligence gathering and management,'' in
  \emph{{SIGMOD}}, 2021.

\bibitem{threatminer}
``{ThreatMiner},'' \url{https://www.threatminer.org/}.

\end{thebibliography}

\end{document}